\documentclass[useAMS,usenatbib]{mn2e}
\usepackage{mn2e-breakabs}
\usepackage{graphicx}
\usepackage{times}
\def\la{\mathrel{\mathpalette\fun <}}
\def\ga{\mathrel{\mathpalette\fun >}}
\def\fun#1#2{\lower3.6pt\vbox{\baselineskip0pt\lineskip.9pt
        \ialign{$\mathsurround=0pt#1\hfill##\hfil$\crcr#2\crcr\sim\crcr}}}

\newcommand{\be}{\begin{equation}}
\newcommand{\ee}{\end{equation}}
\newcommand{\ba}{\begin{eqnarray}}
\newcommand{\ea}{\end{eqnarray}}
\newcommand{\simgt}{\,\hbox{\lower0.6ex\hbox{$\sim$}\llap{\raise0.6ex\hbox{$>$}}}\,}
\newcommand{\simlt}{\,\hbox{\lower0.6ex\hbox{$\sim$}\llap{\raise0.6ex\hbox{$<$}}}\,}

\begin{document}

\title[SDSS DR7 LRG Correlation Function]
{A Robust Distance Measurement and Dark Energy Constraints\\
from the Spherically-Averaged Correlation Function of \\
Sloan Digital Sky Survey Luminous Red Galaxies}

\author[Chuang, Wang, \& Hemantha]{
  \parbox{\textwidth}{
 Chia-Hsun Chuang\thanks{E-mail: chuang@nhn.ou.edu},
 Yun Wang, and
 Maddumage Don P. Hemantha
  }
  \vspace*{4pt} \\
  Homer L. Dodge Department of Physics \& Astronomy, Univ. of Oklahoma,
                 440 W Brooks St., Norman, OK 73019, U.S.A.\\
}

\date{\today}

\maketitle

\begin{abstract}

We measure the effective distance to $z=0.35$, $D_V(0.35)$ 
from the overall shape of the spherically-averaged 
two-point correlation function of the Sloan 
Digital Sky Survey (SDSS) Data Release 7 (DR7) 
luminous red galaxy (LRG) sample. We find $D_V(0.35)=1428_{-73}^{+74}$ 
without assuming a dark energy model or a flat Universe.
We find that the derived measurement of 
$r_s(z_d)/D_V(0.35)=0.1143 \pm 0.0030$ 
(the ratio of the sound horizon at the drag epoch to the effective 
distance to $z=0.35$) is more tightly constrained 
and more robust with respect to possible systematic effects. 
It is also nearly uncorrelated with $\Omega_m h^2$.

Combining our results with the cosmic microwave background and supernova 
data, we obtain $\Omega_k=-0.0032^{+0.0074}_{-0.0072}$ and 
$w=-1.010^{+0.046}_{-0.045}$ (assuming a constant dark energy equation of state).
By scaling the spherically-averaged correlation function, 
we find the Hubble parameter $H(0.35)=83^{+13}_{-15}$ km s$^{-1}$Mpc$^{-1}$ 
and the angular diameter distance $D_A(0.35)=1089^{+93}_{-87}$ Mpc.

We use LasDamas SDSS mock catalogs to compute the covariance matrix 
of the correlation function, and investigate the use of lognormal catalogs
as an alternative. We find that the input correlation function can be accurately
recovered from lognormal catalogs, although they give larger errors on 
all scales (especially on small scales) compared to the mock catalogs
derived from cosmological N-body simulations.

\end{abstract}


\section{Introduction}
Galaxy redshift surveys provide a cosmological probe 
highly complementary to the cosmic microwave
background (CMB) \citep{Penzias:1965wn} and supernovae (SNe)
\citep{Riess:1998cb,Perlmutter:1998np}. 
Large-scale structure data from galaxy surveys can be analyzed using either 
the power spectrum analysis or the correlation function analysis. Although 
these two methods are simple Fourier transforms of one another, the analysis 
processes are quite different and the results cannot be converted with 
Fourier transform directly because of the finite size of the survey volume.
The SDSS data have been analyzed using both the power spectrum
method (see, e.g., \citealt{Tegmark:2003uf,Hutsi:2005qv,Padmanabhan:2006ku,Blake:2006kv,Percival:2007yw,Percival:2009xn,Reid:2009xm}),
and the correlation function method (see, e.g.,
\citealt{Eisenstein:2005su,Okumura:2007br,Cabre:2008sz,Martinez:2008iu,Sanchez:2009jq,Kazin:2009cj}).

The three major uncertainties while constructing a theoretical prediction of the
power spectrum or correlation function with a given cosmological model are the
galaxy bias (the relationship between galaxy and matter distributions), non-linear
effects, and redshift distortions. The knowledge of these uncertainties determines
which analysis method and scale range we should use to obtain reliable
constraints on the dark energy and cosmological parameters.

In this paper, we present the measurement of the spherically-averaged 
correlation function from the SDSS DR7 luminous red galaxy (LRG) 
\citep{Eisenstein:2001cq,Abazajian:2008wr} sample which
provides a homogeneous galaxy sample and has the largest effective survey volume
to date for studying the linear regime \citep{Eisenstein:2005su}. In Section
\ref{sec:data}, we introduce the galaxy sample and selection functions used
in this study. In Section \ref{sec:method}, we describe the details of our
method. In Sec.~\ref{sec:results}, we present our results.
In Sec.~\ref{sec:test}, we check our results using systematic tests.
We summarize and conclude in Sec.~\ref{sec:conclusion}.

\section{Data} \label{sec:data}

The SDSS has observed one-quarter of the
entire sky and performed a redshift survey of galaxies, quasars and
stars in five passbands $u,g,r,i,$ and $z$ with a 2.5m telescope
\citep{Fukugita:1996qt,Gunn:1998vh,Gunn:2006tw}.
We use the public catalog, the NYU Value-Added Galaxy Catalog
(VAGC) \citep{Blanton:2004aa}, derived from the
SDSS II final public data release, Data Release 7 (DR7)
\citep{Abazajian:2008wr}.
We select our LRG sample from the NYU VAGC with
the flag $primTarget$ bit mask set as $32$. K-corrections
have been applied to the galaxies with a
fiducial model ($\Lambda$CDM with $\Omega_m=0.3$ and $h=1$), and
the selected galaxies are required to have rest-frame $g$-band absolute
magnitudes $-23.2<M_g<-21.2$ \citep{Blanton:2006kt}. The same
selection criteria were used in previous work
\citep{Zehavi:2004zn,Eisenstein:2005su,Okumura:2007br,Cabre:2008sz,Kazin:2009cj}.
The sample we use is referred to as ``DR7full'' in \cite{Kazin:2009cj}.
Our sample includes 87000 LRGs in the redshift range 0.16-0.44.
The average weighted redshift of the sample is 0.33.  

Spectra cannot be obtained for objects closer than 55 arcsec
within a single spectroscopic tile due to the finite size of the
fibers. To correct for these ``collisions'', the redshift of an object
that failed to be measured would be assigned to be the same as the
nearest successfully observed one \citep{Zehavi:2004zn}.
Both fiber collision corrections and K-corrections have been made 
in NYU-VAGC \citep{Blanton:2004aa}.

We have applied evolutionary correction based on the stellar
synthesis model of an old and passively evolving burst from
$z=10$ by using the PEGASE code \citep{Fioc:1997sr}.
The rest-frame g-band absolute magnitudes are passively evolved
to $z=0.3$ (see \cite{Zehavi:2004zn,Eisenstein:2005su}).

\begin{figure}
\centering
\includegraphics[width=0.8\columnwidth,clip,angle=-90]{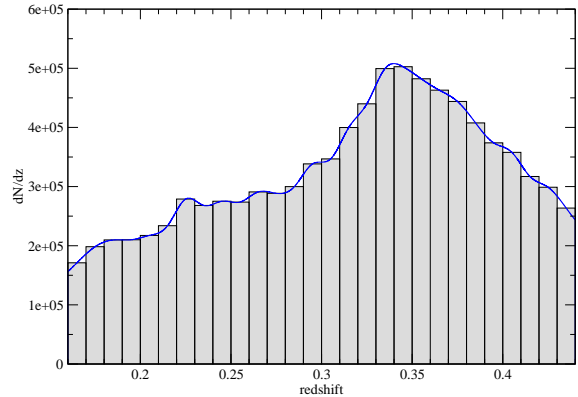}
\caption{The radial selection function of the LRG sample used in this study. 
The gray bars are computed from the sample and 
the blue line is the cubic spline fit of these bar values. We compute 
the radial selection function in the form of the number of galaxies per 
unit redshift instead of the number density in comoving coordinate, so 
that we don't need to assume a fiducial model while generating the random 
catalog with the radial selection function. }
\label{fig:dNdz}
\end{figure}

We construct the radial selection function as a cubic spline fit
to the observed number density histogram with the width $\Delta
z=0.01$ (see Fig.\ \ref{fig:dNdz}). The NYU-VAGC provides the
description of the geometry and completeness of the survey in
terms of spherical polygons. Although the completeness of
VAGC is determined based on the main galaxies
\citep{Strauss:2002dj}, we adopt it as the angular selection
function of our sample since the main galaxies and LRGs should
have similar angular selection functions (see the appendix of
Zehavi et al. 2005). We drop the regions with completeness below
$60\%$ to avoid unobserved plates \citep{Zehavi:2004zn}. The
Southern Galactic Cap region is also dropped.

\section{Methodology}
\label{sec:method}

In this section, we describe the measurement of the correlation function
from the observational data, construction of the theoretical prediction,
and the likelihood analysis that leads to constraints on
dark energy and cosmological parameters.
We will also show that using one scaling parameter is sufficient for
extracting information from the observed spherically
averaged correlation function (see Sec.~\ref{sec:proof}).

\subsection{Measuring the Two-point Correlation Function}
\begin{figure}
\centering
\includegraphics[width=0.8\columnwidth,clip,angle=-90]{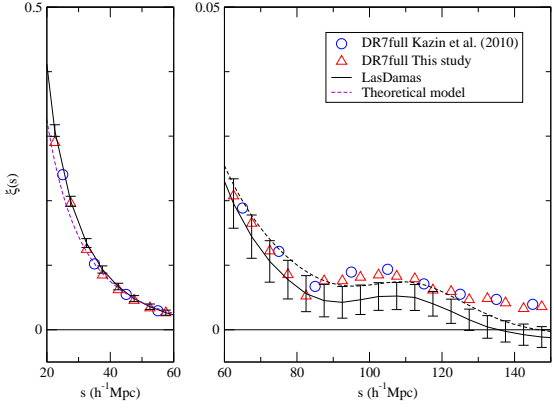}
\caption{The spherically-averaged two-point correlation
function measured from the SDSS DR7 data. The red triangles are the correlation
function computed with the LRG sample described in Sec.~\ref{sec:data}.
The green circles are taken from \citealt{Kazin:2009cj} in which
the same fiducial model is used ($\Lambda$CDM with $\Omega_m=0.25$)
but the bin size they use is $10h^{-1}$Mpc.
Our result shows excellent agreement with that of \citealt{Kazin:2009cj}.
The black line is the average correlation function from LasDamas mock catalogs.
The error bars are the square roots of the diagonal elements of the covariance matrix 
we have derived (see Sec.~\ref{sec:covar}). 
The violet dashed line is the mean model from our MCMC likelihood
analysis ($\Omega_m h^2=0.105, \Omega_b h^2=0.0225, n_s=0.978, D_V(0.35)=1428$Mpc).
Note that an MCMC analysis does not result in an accurate bestfit model 
\citep{Lewis:2002ah}.}
\label{fig:kazin_my}
\end{figure}

We calculate the comoving distances to every galaxy by assuming a
fiducial model, $\Lambda$CDM with $\Omega_m=0.25$. We use the
two-point correlation function estimator given by Landy and Szalay
\citep{Landy:1993yu}:

\begin{equation}
\xi(s) = \frac{DD(s)-2DR(s)+RR(s)}{RR(s)},
\end{equation}
where DD, DR, and RR represent the normalized data-data,
data-random, and random-random pair counts respectively in a
distance range. The bin size we use in this study is
$5h^{-1}$Mpc. This estimator has minimal variance for a Poisson
process. Random data should be generated according to the radial
and angular selection functions of the data. One can reduce 
the shot noise due to random data by increasing the number of 
random data. The number of random data we use is 10 times that 
of the real data. While calculating the pair counts, we assign 
each data point a radial weight of $1/[1+n(z)\cdot P_w]$, where 
$n(z)$ is the radial selection function and $P_w = 4\cdot 10^4$ 
$h^{-3}$Mpc$^3$ as in \cite{Eisenstein:2005su}. The observed 
correlation function is shown in Fig.~\ref{fig:kazin_my}.

\subsection{Theoretical Two-Point Correlation Function}
We compute the linear power spectra at $z=0.33$ by using CAMB
\citep{Lewis:1999bs}. To include the damping effect of non-linear structure
formation and peculiar velocities, we calculate the
dewiggled power spectrum

\begin{equation} \label{eq:dewiggle}
P_{dw}(k)=P_{lin}(k)\exp\left(-\frac{k^2}{2k_\star^2}\right)
+P_{nw}(k)\left[1-\exp\left(-\frac{k^2}{2k_\star^2}\right)\right],
\end{equation}
where $P_{lin}(k)$ is the linear power spectrum, $P_{nw}(k)$ is the no-wiggle 
or pure CDM power spectrum calculated with Eq.(29) in \cite{Eisenstein:1997ik}, 
and $k_{\star}$ is marginalized over with a flat prior over the range of
0.09 to 0.13\footnote{Although $k_{\star}$ can be computed by 
renormalization perturbation theory \citep{Crocce:2005xz,Matsubara:2007wj}, 
doing so requires knowing the amplitude of the power spectrum, which is 
also marginalized over in this study.}

\begin{figure}
\centering
\includegraphics[width=0.8\linewidth,clip,angle=-90]{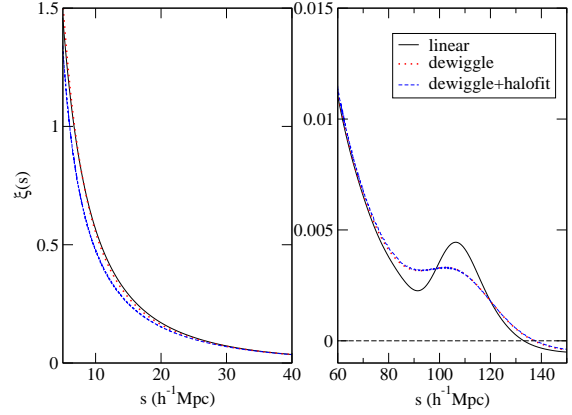}
\caption{An example of the effect of applying dewiggle and \emph{halofit} to
the correlation function. The black solid line is the linear correlation function
without applying dewiggle and \emph{halofit} yet. The red dotted line is the
dewiggled linear correlation function.
The green dashed line is the dewiggled correlation function including
nonlinear effects calculated using \emph{halofit}.
The damping of BAO is accurately described by the dewiggled linear correlation function.
Additional nonlinear effects are only important on very small scales. }
  \label{fig:dw_halofit}
\end{figure}

We then use the software package \emph{halofit} \citep{Smith:2002dz} to compute the 
non-linear matter power spectrum:
\begin{eqnarray} \label{eq:halofit}
r_{\emph{halofit}}(k) &\equiv& \frac{P_{\emph{halofit},nw}(k)}{P_{nw}(k)} \\
P_{nl}(k)&=&P_{dw}(k)r_\emph{halofit}(k),
\end{eqnarray}
where $P_{\emph{halofit},nw}(k)$ is the power spectrum from applying \emph{halofit} 
on the no-wiggle power spectrum and $P_{nl}(k)$ is the non-linear power spectrum.
We compute the theoretical two-point correlation
function by Fourier transforming the non-linear power spectrum. We show
an example of the effect of applying dewiggle and \emph{halofit} to the
correlation function in Fig.\ \ref{fig:dw_halofit}. Clearly,
the damping of the baryon acoustic oscillation (BAO) peak 
is accurately described by the dewiggled linear
correlation function. Additional nonlinear effects are only important
on very small scales.

The parameter set we use to compute the theoretical
correlation function is
$\{D_V(z), \Omega_mh^2, \Omega_bh^2, n_s, k_\star\}$, where
$\Omega_m$ and $\Omega_b$ are the density fractions of matter and
baryons, $n_s$ is the power law index of the primordial matter power spectrum,
$h$ is the dimensionless Hubble
constant ($H_0=100h$ km s$^{-1}$Mpc$^{-1}$), and  $D_V(z)$ is defined by
\begin{equation} \label{eq:dv}
 D_V(z)\equiv \left[(1+z)^2D_A^2\frac{cz}{H(z)}\right]^\frac{1}{3},
\end{equation}
where $H(z)$ and $D_A(z)$ are the Hubble parameter and the angular diameter 
distance at the redshift, $z$. We set $h=0.7$ while calculating the non-linear 
power spectra. Assuming the true model of our universe is not far from a $\Lambda$CDM model, 
the dark energy and curvature dependence are absorbed by the
effective distance, $D_V(z)$. Thus we are able to extract
constraints from data without assuming a dark energy model and
cosmic curvature.

\subsection{Covariance Matrix} \label{sec:covar}

We use the mock catalogs from the LasDamas
simulations\footnote{http://lss.phy.vanderbilt.edu/lasdamas/}
(McBride et al., in preparation)
to estimate the covariance matrix of the observed correlation function.
LasDamas provides mock catalogs matching SDSS main galaxy and LRG samples.
We use the LRG mock catalogs from the LasDamas gamma release with the same cuts as
the SDSS LRGfull sample, $-23.2<M_g<-21.2$ and $0.16<z<0.44$.
We have diluted the mock catalogs to match the radial selection function
of the observed data by randomly selecting the mock galaxies according to the
number density of the data sample. 
We calculate the spherical-averaged correlation functions
of the mock catalogs and construct the covariance matrix as

\begin{equation}
 C_{ij}=\frac{1}{N-1}\sum^N_{k=1}(\bar{\xi}_i-\xi_i^k)(\bar{\xi}_j-\xi_j^k),
\label{eq:covmat}
\end{equation}
where $N$ is the number of the mock catalogs, $\bar{\xi}_m$ is the
mean of the $m^{th}$ bin of the mock correlation functions, and
$\xi_m^k$ is the value of $m^{th}$ bin of the $k^{th}$ mock
correlation function. 
We test the normality of the correlation functions from the LasDamas mock 
catalogs and find that they are well described by normal distributions 
(see Appendix\ \ref{sec:normality_test}).

\begin{figure}
\centering
\includegraphics[width=0.8\linewidth,clip,angle=-90]{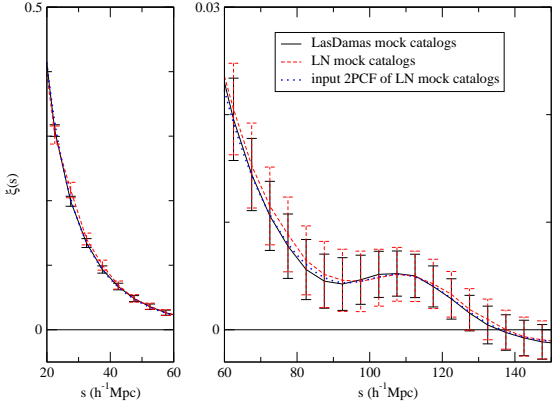}
\caption{Spherical-averaged 2PCF of the mock catalogs. The black solid line 
is computed from the LasDamas mock catalogs. The red dashed line is computed 
from our lognormal mock catalogs. The error bars are the square roots 
of the dianogal elements of the covariance matrixes. The green dotted line 
is the input 2PCF for our lognormal mock catalogs.}
  \label{fig:compare_lasdamas_ln}
\end{figure}

\begin{figure}
\centering
\includegraphics[width=1\linewidth,clip,angle=0]{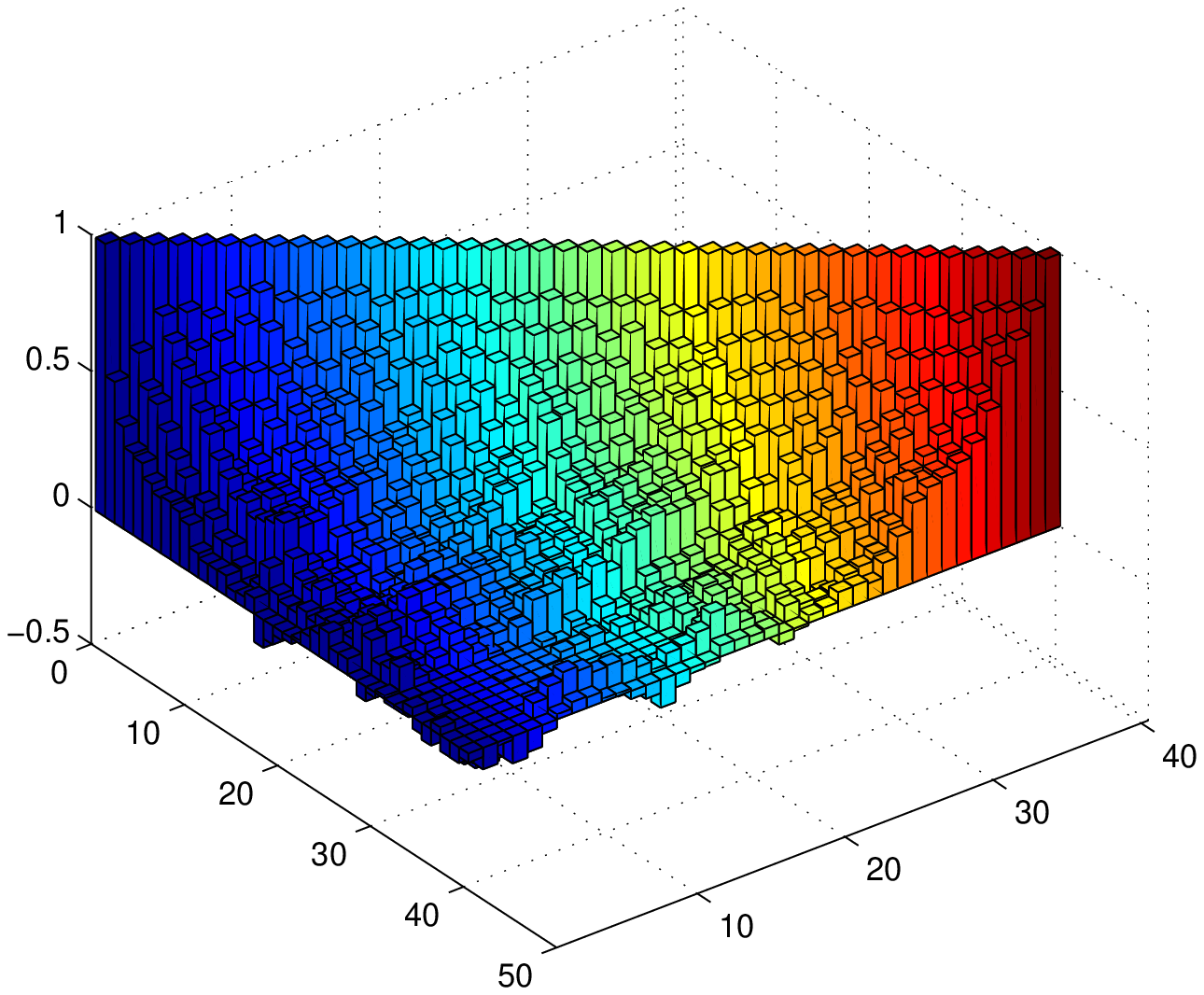}
\caption{The normalized covariance matrix computed from 160 LasDamas mock catalogs. We show the covariance among 40 bins from $0<s<200 h^{-1}$Mpc with bin size $5h^{-1}$Mpc.}
  \label{fig:lasdamas_covar}
\centering
\includegraphics[width=1\linewidth,clip,angle=0]{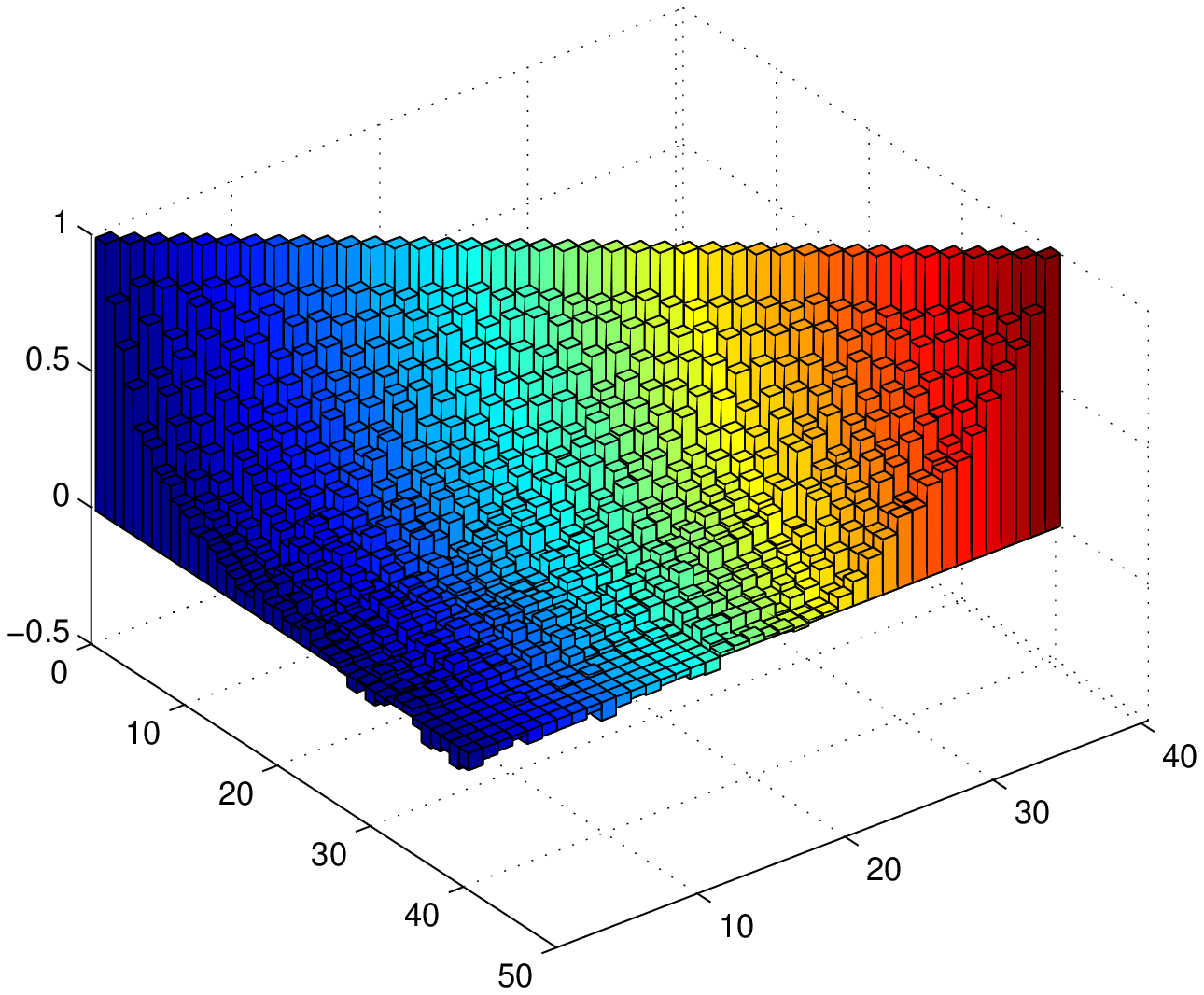}
\caption{The normalized covariance matrix computed from 500 lognormal mock catalogs. We show the covariance among 40 bins between the scale range, $0<s<200 h^{-1}$Mpc, with the bin size, $5h^{-1}$Mpc. }
  \label{fig:ln_covar}
\end{figure}

\begin{figure}
\centering
\includegraphics[width=0.8\linewidth,clip,angle=-90]{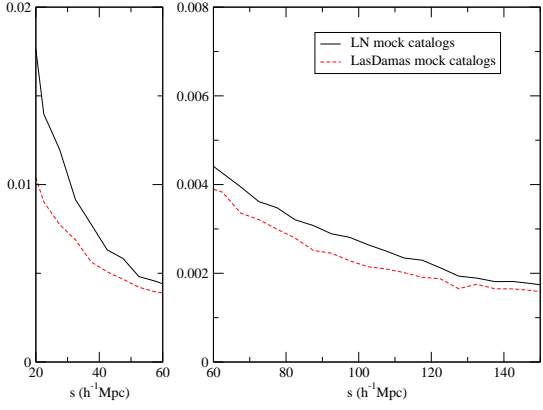}
\caption{Square roots of the dianogal elements of the covariance matrixes. The black dashed line is computed from the LasDamas mock catalogs. The red solid line is computed from lognormal mock catalogs we create. One can see that LN mock catalogs have larger covariance at smaller scale and two line are close at larger scale.}
  \label{fig:compare_errors}
\end{figure}

The mock catalogs derived from N-body simulations require long computing times
and are very limited in availability. It is interesting to investigate 
whether there is an easier, faster, and cheaper way to construct mock catalogs 
which could work as well as those derived from N-boday simulation. 
Towards this end, we have created 500 lognormal(LN) mock catalogs 
\citep{Coles:1991if,Percival:2003pi}, and computed the spherically-averaged 
correlation functions from these. The details involved in creating LN mock 
catalogs are described in Appendix\ \ref{sec:lognormal}.
We compare the correlation funcions from the LasDamas mock catalogs and
LN mock catalogs in Fig.\ \ref{fig:compare_lasdamas_ln}; the error bars
indicate the square roots of the dianogal elements of the covariance matrixes. 
We also show the normalized covariance matrixes in Fig.\ \ref{fig:lasdamas_covar} 
and \ref{fig:ln_covar}. Clearly, the results from the LasDamas mocks and our LN mocks
are very similar to each other. In particular, the input correlation function
is accurately recovered by analyzing the LN mock catalogs. Note that
the LN mocks give larger errors on all scales, and on scales
smaller than $\sim 40 h^{-1}$Mpc, the LN mock catalogs give much larger
errors than the LasDamas mock catalogs (see Fig.\ \ref{fig:compare_errors}).

We use the covariance matrix computed from the LasDamas SDSS mock catalogs, 
since these are more realistic than the lognormal mock catalogs, 
and give smaller errors for the measured correlation function.
It is interesting to note that in the absence of mock catalogs
derived from cosmological N-body simulations, lognormal catalogs
can be used for a conservative estimate of the covariance matrix
of the correlation function.

\subsection{Likelihood}
The likelihood is taken to be proportional to $\exp(-\chi^2/2)$, and $\chi^2$ is given by

\begin{equation} \label{eq:chi2}
 \chi^2\equiv\sum_{i,j=1}^{N_{bins}}\left[\xi_{th}(s_i)-\xi_{obs}(s_i)\right]
 C_{ij}^{-1}
 \left[\xi_{th}(s_j)-\xi_{obs}(s_j)\right]
\end{equation}
where $N_{bins}$ is the number of bins used, $\xi_{th}$ is the theoretical
 correlation function of a model, and $\xi_{obs}$ is the observed correlation function.
 Note that $\xi_{th}(s_i)$ depends
on $\{D_V(z), \Omega_mh^2, \Omega_bh^2, n_s, k_{\star}\}$.

In principle, we should recalculate the observed correlation function while
computing the $\chi^2$ for different models. However, since we don't consider
the entire scale range of the correlation function (we only consider
$s=40-120\ h^{-1}$Mpc in this study), we might include or exclude different
data pairs for different models which would render $\chi^2$ values arbitrary.
Therefore, instead of recalculating the observed correlation function, we
apply the inverse operation to the theoretical correlation function
to move the parameter dependence from the data to the model,
thus preserving the number of galaxy pairs used in the likelihood analysis.

Let us define $T$ as the operator converting the
measured correlation function from the fiducial model to another
model, i.e.,
\begin{equation}
 \xi_{obs}(s)=T(\xi_{obs}^{fid}(s)).
\label{eq:T def}
\end{equation}
where $\xi^{fid}_{obs}(s)$ is the observed correlation function
assuming the fiducial model. This allows us to rewrite $\chi^2$ as
\ba
\label{eq:chi2_2}
\chi^2 &\equiv&\sum_{i,j=1}^{N_{bins}}
 \left\{T^{-1}\left[\xi_{th}(s_i)\right]-\xi^{fid}_{obs}(s_i)\right\}
 C_{fid,ij}^{-1} \cdot \nonumber\\
 & & \cdot \left\{T^{-1}\left[\xi_{th}(s_j)\right]-\xi_{obs}^{fid}(s_j)\right\},
\ea
where we have used Eqs.(\ref{eq:covmat}) and (\ref{eq:T def}).

To find the operator $T$, note that the fiducial model is only used
in converting redshifts into distances for the galaxies in our data
sample. In the analysis of galaxy clustering, we only need the
separation of a galaxy pair, and not the absolute distances to
the galaxies. For a thin redshift shell, we can convert
the separation of one pair of galaxies from the fiducial model to
another model by performing the scaling
(see, e.g., \cite{SE03})
\begin{equation}
 s'=\sqrt{\left(s\cos\theta\frac{H^{fid}(z)}{H(z)}\right)^2+
 \left(s\sin\theta\frac{D_A(z)}{D_A^{fid}(z)}\right)^2},
\label{eq:convert_s_2d}
\end{equation}
where $\theta$ is the angle between the radial direction and the direction of the
line connecting the pair of galaxies.

\cite{Eisenstein:2005su} argued that we can use one rescaling parameter, 
$D_V(z)$, to convert the observed correlation function from the fiducial 
model to another model as long as the new model is not very different from 
the fiducial one, and the redshift range of the sample is not large.
Then the separation of one pair of galaxies is converted from the fiducial model 
to another by
\begin{equation}
 s'=\frac{D_V(z)}{D_V^{fid}(z)}s.
\label{eq:convert_s_1d}
\end{equation}
In this section, we discuss methods with one and two rescaling 
parameters, and show that these two methods are equivalent 
for spherically-averaged data when certain conditions hold (see Sec.\ \ref{sec:proof}).

\subsubsection{Using One Rescaling Parameter} \label{sec:one_rescaling}

From eq.\ (\ref{eq:convert_s_1d}), the observed correlation function with the 
different model can be written as follows:
\begin{equation} \label{eq:xi_obs}
 \xi_{obs}(s)=\xi^{fid}_{obs}\left(\frac{D_{V}^{fid}(z_{eff})}{D_{V}(z_{eff})}\, s\right),
\end{equation}
where $z_{eff}$ is the effective redshift of the sample and $D_V(z)$ is 
defined by Eq.(\ref{eq:dv}).

The effective redshift we use in this study is $z_{eff}=0.33$.
Since the results are insensitive to $z_{eff}$ (see Sec.\ \ref{sec:test}), we rescale our result to $z_{eff}=0.35$ for comparing with previous works.
Eq.\ (\ref{eq:xi_obs}) can be rewritten as
\begin{equation} \label{eq:xi_fid}
 \xi^{fid}_{obs}(s)=T^{-1}(\xi_{obs}(s))=\xi_{obs}\left(\frac{D_{V}(z_{eff})}
 {D_{V}^{fid}(z_{eff})} s\right).
\end{equation}

We can apply the same inverse rescaling operation to the theoretical correlation function:
\begin{equation} \label{eq:inverse_theory}
 T^{-1}(\xi_{th}(s))=\xi_{th}\left(\frac{D_{V}(z_{eff})}{D_{V}^{fid}(z_{eff})}s\right).
\end{equation}
$\chi^2$ can be calculated by substituting eq.\ (\ref{eq:inverse_theory}) into
eq.\ (\ref{eq:chi2_2}).

\subsubsection{Using Two Rescaling Parameters} 
\label{sec:two_rescaling}

From eq.\ (\ref{eq:convert_s_2d}), we can convert the 
spherically-averaged correlation function from some model to the fiducial model by
\begin{eqnarray}
 \xi^{fid}_{obs}(s)&=&T^{-1}(\xi_{obs}(s))\nonumber\\
 &=&\int_0^\pi d\theta w(s,\theta) \times \nonumber
\end{eqnarray}
\begin{equation}
 \xi_{obs}\left(\sqrt{\left(s\cos\theta\frac{H^{fid}(z)}{H(z)}\right)^2
 +\left(s\sin\theta\frac{D_A(z)}{D_A^{fid}(z)}\right)^2}\right),
\end{equation}
where the weighting function $w(r,\theta)$ is given by
\begin{equation}
 w(s,\theta)=\frac{n_{DD}(s,\theta)}{\int_0^\pi d\theta n_{DD}(s,\theta)},
\end{equation}
where $n_{DD}(s,\theta)$ is the number density of the data pairs. We define
inverse operation, $T^{-1}$, directly since $T$ is not necessary in our calculation.
We now apply the inverse operation to the theoretical correlation function:
\begin{eqnarray}
 T^{-1}(\xi_{th}(s))=\int_0^\pi d\theta w(s,\theta) \times \nonumber
\end{eqnarray}
\begin{equation}\label{eq:inverse_theory_2d}
 \xi_{th}
 \left(\sqrt{\left[s\cos\theta\frac{H^{fid}(z)}{H(z)}\right]^2+
 \left[s\sin\theta\frac{D_A(z)}{D_A^{fid}(z)}\right]^2}\right).
\end{equation}
$\chi^2$ can be calculated by substituting eq.\ (\ref{eq:inverse_theory_2d})
into eq.\ (\ref{eq:chi2_2}).

\subsubsection{Equivalence of Using One and Two Rescaling Parameters
for Spherically-Averaged Data} \label{sec:proof}

We now show that using one and two rescaling parameters while calculating
the spherically-averaged correlation function are equivalent to first order
in approximation. If the size of the survey is much larger than the scales of 
interest, $n_{DD}(s,\theta)$ would be proportional to $s\sin\theta$.
Hence
\begin{equation}
 w(s,\theta)\sim\frac{s\sin\theta}{\int_0^\pi s\sin\theta d\theta}=\frac{\sin\theta}{2}.
\end{equation}
Next, if the model is close to the fiducial model, we can just consider the first order
terms of $D_V/D_V^{fid}$, $H^{fid}/H$, and $D_A/D_A^{fid}$ which can be written as following:
\begin{equation}
 \frac{D_V}{D_V^{fid}}\simeq 1+\delta_V,\  \frac{H^{fid}}{H}\simeq 1+\delta_r,\
 \frac{D_A}{D_A^{fid}}\simeq 1+\delta_a,
\end{equation}
where $|\delta_V|, |\delta_r|, |\delta_a| \ll 1$. From the definition of $D_V$
(see Eq.[\ref{eq:dv}]), one can obtain a simple relation, $3\delta_V\simeq\delta_r+2\delta_a$.
Let's consider a power law correlation function:
\begin{equation}
 \xi_{th}(s)=s^p,
\end{equation}
where p is a real number. Eq.(\ref{eq:inverse_theory_2d}) can be rewritten as
\ba\label{eq:xi_theory_2}
 &&T^{-1}(\xi_{th}(s)) \nonumber\\
 &\simeq&\int_0^\pi d\theta \frac{\sin\theta}{2}
 \left(\sqrt{\left[s\cos\theta(1+\delta_r)\right]^2
 +\left[s\sin\theta(1+\delta_a)\right]^2}\right)^p\nonumber\\
\nonumber &\simeq& \frac{s^p}{2} \int_0^\pi d\theta \sin\theta
\left[\cos^2\theta(1+2\delta_r)+\sin^2\theta(1+2\delta_a)\right]^\frac{p}{2}\\
\nonumber &\simeq& \frac{s^p}{2} \int_0^\pi d\theta \sin\theta
\left[1+p(\cos^2\theta \delta_r+\sin^2\theta \delta_a)\right]\\
\nonumber &=& s^p\left[1+p\frac{\delta_r+2\delta_a}{3}\right]\\
\nonumber &\simeq& s^p(1+p\delta_V)\\
\nonumber &\simeq& \left(\frac{D_V}{D_V^{fid}}s\right)^p\\
 &=&\xi_{th}\left(\frac{D_V}{D_V^{fid}} s\right)
\ea
The proof can be generalized to any function which can be expressed as
\begin{equation}
 \xi_{th}(s)=s^{p_1}+s^{p_2}+s^{p_3}+...
\end{equation}
where $p_i$ are real numbers.

To measure the spherically-averaged correlation function, we have shown that
using one rescaling parameter, $D_V$, and two rescaling parameters, $H$ and $D_A$,
are equivalent as long as the scales of interest are relatively small compared to
the survey length scale, and the constraint on $D_V$ is tight enough. A similar
statement can be made for the spherically-averaged power spectrum analysis.
We have verified that these two rescaling methods give similar results.

\subsection{Markov Chain Monte-Carlo Likelihood Analysis}

We use CosmoMC \citep{Lewis:2002ah} in a Markov Chain Monte-Carlo
likelihood analysis. The main parameter space that we explore is
$\{\Omega_mh^2, \Omega_bh^2, n_s, D_V(z_{eff}), k_\star \}$ and the prior ranges are
$\{(0.025,0.3)$, $(0.01859,0.02657)$, $(0.865,1.059)$, $(725,1345)$, $(0.09,0.13)\}$ respectively.
The dependence on $h$, the curvature, and dark energy
parameters are absorbed into $D_V(z_{eff})$.

We marginalize over the amplitude of the correlation function;
this is equivalent to marginalizing over galaxy bias$\times \sigma_8 \times r_\beta$, 
where $\sigma_8$ is the matter power spectrum normalization parameter and $r_\beta$ 
is the linear ratio between the correlation function in the redshift space and 
real space which can be derived from the linear redshift distortion parameter 
\citep{Kaiser:1987qv}.
Since the LRG data alone cannot give tight constraints on $\Omega_b h^2$ and $n_s$, 
we apply flat priors ($\pm 7\sigma_{WMAP}$) on them which are wide enough so that CMB
constraints will not be double counted. In other words, the effect from the wide 
flat priors could be ignored when combining our final results with CMB data.
We also marginalized over $k_*$ over the range of 0.09 to 0.13
(see Sec.3.2).

\section{Results} \label{sec:results}
In this section, we present the model independent measurements of the
parameters we explore, $\{D_V(0.35)$, $\Omega_m h^2\}$, obtained by using the method described in previous sections.
Although, the effective redshift we use is 0.33, the average weighted redshift, we rescale all our results to $z_{eff}=0.35$ for comparing with previous work easily by
\be
 D_V(0.35)=D_V(0.33)\frac{D_{V,fid}(0.35)}{D_{V,fid}(0.33)}=1.054 D_V(0.33).
\ee
We have checked that the results is insensitive to the effective redshift in Sec.\ \ref{sec:test}.
 
We derive the model independent measurements of $H$ and $D_A$ for comparison with 2D results. We also apply the method on two subsamples (two redshift slices) as a systematic test.  

We validate our method by applying it to the LasDamas mock catalogs,
and find that our measurements are consistent with the input
parameters of the simulations.

We derive constraints on dark energy and cosmological
parameters by combining our results with other data sets including
WMAP7 \citep{Komatsu:2010fb} and Union2 SN
\citep{Amanullah:2010vv}.

Finally, we compare our results with previous works.

\subsection{Model Independent Constraints on $D_V(0.35)$} \label{sec:no-model}

Without assuming a dark energy model or a flat Universe,
we find that $D_V(0.35)=1428_{-73}^{+74}$ Mpc 
and $r_s(z_d)/D_V(0.35) =0.1143 \pm 0.0030$, 
where $r_s(z_d)$ is the comoving sound horizon at the drag epoch calculated
with the eq.\ (6) in \cite{Eisenstein:1997ik}. 
 
\begin{figure}
\centering
\includegraphics[width=0.8\linewidth,clip]
{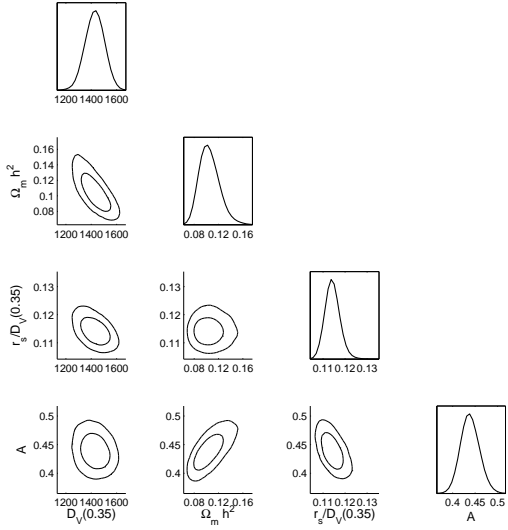}
\caption{2D marginalized
contours ($68\%$ and $95\%$ C.L.) for $D_V(0.35)$, $\Omega_m h^2$, $r_s(z_d)/D_V(0.35)$, and $A(0.35)$. The diagonal
  panels represent the marginalized probabilities.}
  \label{fig:dv5params}
\end{figure}

Fig.\ \ref{fig:dv5params} shows one and two-dimensional marginalized
contours of the parameters, $\{D_V(0.35)$, $\Omega_m h^2$, $r_s(z_d)/D_V(0.35)$, $A(0.35)\}$, where
\begin{equation}\label{eq:A}
 A(z)\equiv D_V(z)\frac{\sqrt{\Omega_mH_0^2}}{cz}.
\end{equation}
 The measurements and the covariance
matrix are listed in Table \ref{table:mean} and
\ref{table:covar_matrix}. The bestfit model from the MCMC
likelihood analysis has $\chi^2=6.32$ for 16 bins of data used (in
the scale range of $40 \,h^{-1}$Mpc$<s<120\,h^{-1}$Mpc with the
bin size = $5\,h^{-1}$Mpc), for a set of 6 parameters (including
the overall amplitude of the correlation function).

The scale range of the correlation function we have selected
is $s=40-120\ h^{-1}$Mpc. In this range, the scale dependence of
the redshift distortion and galaxy bias is small. We cut the tail of the correlation
function at $s=120\ h^{-1}$Mpc because the high tail (large correlation at
large scales) cannot be fitted to any conventional model, and could be due to
systematic error or sample variance
(see further discussion in Sec.\ \ref{sec:test}).

At this point, we assume the high tail is simply due to
sample variance, and might disappear when much larger data sets become available.
Unlike previous analyses by other groups, we apply very weak
flat priors ($\pm 7\sigma_{WMAP7}$) on $\Omega_b h^2$ and $n_s$ instead of fixing them
to the bestfit values from CMB data.

\begin{table}
\begin{center}
\begin{tabular}{crrrr}\hline
&mean &$\sigma$ &lower &upper \\ \hline
$D_V(0.35)$(Mpc)&\ \  1428&\ \  74&\ \  1355&\ \  1502 \\
$\Omega_mh^2$&\ \ 0.105&\ \  0.016&\ \  0.090&\ \  0.121\\
$r_s(z_d)/D_V(0.35)$&\ \ \bf{0.1143}&\ \ \bf{0.0033}&\ \  \bf{0.1113}&\ \  \bf{0.1173}\\
$A(0.35)$&\ \ 0.439&\ \  0.020&\ \  0.419&\ \ 0.459 \\ \hline
\hline
\end{tabular}
\end{center}
\caption{Measured cosmological parameters with flat prior $0.01859<\Omega_bh^2<0.02657$,
$0.865<n_s<1.059$ ($\pm 7\sigma_{WMAP7}$), and $0.09<k_*<0.13 h^{-1}/$Mpc. The standard deviations and the marginalized
bounds (68\%) are listed as well. There are two derived measurements, $r_s(z_d)/D_V(0.35)$ and $A(0.35)$, in the table.
} \label{table:mean}
\end{table}

\begin{table}
\begin{center}
\begin{tabular}{crrrr}\hline
       &$D_V(0.35)$  &$\Omega_mh^2$&  $\frac{r_s(z_d)}{D_V(0.35)}$ &$A(0.35)$\\ \hline
$D_V(0.35)$&\ \  1&\ \ -0.7890&\ \ -0.5561&\ \ -0.1727\\
$\Omega_mh^2$&\ \ -0.7890&\ \ 1&\ \ 0.0056&\ \ 0.7305\\
$\frac{r_s(z_d)}{D_V(0.35)}$ &\ \ -0.5561&\ \ 00056&\ \ 1&\ \ -0.6181\\
$A(0.35)$&\ \ -0.1727&\ \  0.7305&\ \  -0.6181&\ \ 1 \\
\hline
\end{tabular}
\end{center}
\caption{Normalized covariance matrix with flat prior
$0.01859<\Omega_bh^2<0.02657$, $0.865<n_s<1.059$ ($\pm 7\sigma_{WMAP7}$), and $0.09<k_*<0.13 h^{-1}/$Mpc.}
 \label{table:covar_matrix}
\end{table}

\subsection{Model independent measurements of $H(0.35)$ and $D_A(0.35)$} \label{sec:hda}

\begin{figure}
\centering
\includegraphics[width=0.8\linewidth,clip]
{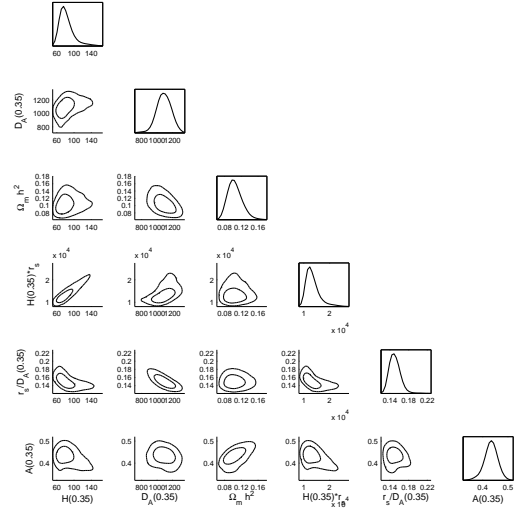}
\caption{2D marginalized
  contours for $68\%$ and $95\%$ for $H(z=0.35)$, $D_A(z=0.35)$,
  $\Omega_m h^2$, $H(0.35)*r_s(z_d)$, $r_s(z_d)/D_A(0.35)$, and $A(0.35)$. The diagonal
  panels represent the marginalized probabilities.}
  \label{fig:hda5params}
\end{figure}

In this section, we apply the method with two scaling parameters described
in sec \ref{sec:two_rescaling} to measure $H$ and $D_A$. 
The main parameter space that we explore is
$\{\Omega_mh^2, \Omega_bh^2, n_s, H(z_{eff}), D_A(z_{eff}), k_\star \}$ and the prior ranges are
$\{(0.025,0.3)$, $(0.01859,0.02657)$, $(0.865,1.059)$, $(41,123)$, $(723,1343)$, $(0.09,0.13)\}$ respectively.
We obtain the model
independent measurements, $H(0.35)=83^{+13}_{-15}$ km s$^{-1}$Mpc$^{-1}$
and $D_A(0.35)=1089^{+93}_{-87}$ Mpc, from the LRG data alone (see
Table\ \ref{table:mean_hda}). Table\ \ref{table:covar_matrix_hda} shows
the normalized covariance matrix of $\{H(0.35)$,
$D_A(0.35)$, $\Omega_mh^2$, $H(0.35)*r_s(z_d)$, $r_s(z_d)/D_A(0.35)$,$A(0.35)\}$, and
Fig.\ \ref{fig:hda5params} shows the 2D marginalized contours of this
parameter set.

Although using two rescaling parameters on the spherically-averaged
correlation function cannot give better constraints on the cosmological
parameters, it gives the model independent measurements of $H$ and
$D_A$ which cannot be derived directly from the measurement of $D_V$. 
These can be compared to our result for the 
two-dimensional two-point correlation function \citep{Chuang:2011fy}, 
$H(0.35)=82.1_{-4.9}^{+4.8}\,{\rm km}\,{\rm s}^{-1}\,{\rm Mpc}^{-1}$ and
$D_A(0.35)=1048_{-58}^{+60}$ Mpc.
Not surprisingly, information is lost in the spherical averaging of data.

\begin{table}
\begin{center}
\begin{tabular}{crrrr}\hline
&mean &$\sigma$ &lower &upper \\ \hline
$H(0.35)$ (km s$^{-1}$Mpc$^{-1}$)&\ \  83&\ \   17&\ \  68&\ \  96\\
$D_A(0.35)$ (Mpc)&\ \  1089&\ \  93&\ \  1002&\ \  1182\\
$\Omega_mh^2$&\ \ 0.105&\ \  0.017&\ \  0.089&\ \  0.122\\
$H(0.35)*r_s(z_d)$&\ \ 13500&\ \ 2700 &\ \  11200&\ \  15500\\
$r_s(z_d)/D_A(0.35)$&\ \ 0.151&\ \  0.012&\ \  0.140&\ \  0.161 \\
$A(0.35)$&\ \ 0.432&\ \  0.026&\ \  0.408&\ \  0.457 \\ \hline
\end{tabular}
\end{center}
\caption{Measured cosmological parameters with flat prior
$0.01859<\Omega_bh^2<0.02657$,
$0.865<n_s<1.059$ ($\pm 7\sigma_{WMAP7}$), and $0.09<k_*<0.13 h^{-1}/$Mpc. The standard
deviations and the margenalized bounds (68\%) are listed as well. There are three derived measurements, $H(0.35)*r_s(z_d)$, $r_s(z_d)/D_A(0.35)$, and $A(0.35)$, in the table.
} \label{table:mean_hda}
\end{table}

\begin{table*}
\begin{center}
\begin{tabular}{crrrrrr}\hline
       &$H(0.35)$  &$D_A(0.35)$  &$\Omega_mh^2$&  $H(0.35)*r_s(z_d)$ &$r_s(z_d)/D_A(0.35)$&$A(0.35)$\\ \hline
$H(0.35)$&\ \      1&\ \    0.4028&\ \    0.2133&\ \   0.9744&\ \   -0.5733&\ \ -0.3460\\
$D_A(0.35)$&\ \      0.4028&\ \    1&\ \   -0.4829&\ \  0.5300&\ \   -0.8440&\ \ -0.1210\\
$\Omega_mh^2$&\ \    0.2133&\ \   -0.4829 &\ \   1&\ \  0.0025&\ \   -0.0139&\ \ 0.6035\\
$H(0.35)*r_s(z_d)$&\ \     0.9744&\ \  0.5300&\ \    0.0025&\ \    1&\ \   -0.5861&\ \ -0.4723\\
$r_s(z_d)/D_A(0.35)$&\ \    -0.5733&\ \   -0.8440&\ \   -0.0139&\ \    -0.5861&\ \    1&\ \ -0.2033\\
$A(0.35)$&\ \    -0.3460&\ \    0.0908&\ \ 0.6035&\ \    -0.4723&\ \    -0.2033&\ \ 1 \\ \hline
\end{tabular}
\end{center}
\caption{Normalized covariance matrix with flat prior
$0.01859<\Omega_bh^2<0.02657$, $0.865<n_s<1.059$ ($\pm 7\sigma_{WMAP7}$), and $0.09<k_*<0.13 h^{-1}/$Mpc.
} \label{table:covar_matrix_hda}
\end{table*}

\subsection{Model independent measurements of $r_s(z_d)/D_V(z)$ from two redshift slices} \label{sec:2slice}
To explore the redshfit dependency of the measurements. We apply the method of one rescaling parameter on two subsamples have $z=0.16-0.36$ and $z=0.28-0.44$. The average weighted redshifts of these two samples are 0.28 and 0.36 respectively. Both subsamples have $\sim 60000$ galaxies. We find $r_s(z_d)/D_V(0.28)=0.141 \pm 0.012$ and $r_s(z_d)/D_V(0.36)=0.1146 \pm 0.0068$. Because of the overlapping of the samples, there is covariance between these two measurements. To estimate the covariance, we apply the method on 40 mock catalogs (indexed from 01a to 40a) and find the correlation coefficient r = 0.20. To combine these measurements with other data sets, one should add the following term to the $\chi^2$:
\be \label{eq:chi2_2slice}
\chi^2_{LRG2Z}=
\Delta_{LRG2Z}
\left(\begin{array}{cc}6688 & -2437 \\ -2437 & 22200\end{array}\right)
\Delta_{LRG2Z}
\ee
where 
\be
 \Delta_{LRG2Z}=\left(\begin{array}{c}r_s(z_d)/D_A(0.28)-0.1413\\ r_s(z_d)/D_A(0.36)-0.1146\end{array}\right)
\ee

The constraints of cosmological parameters from combining CMB and SNe data set 
for the owCDM model are shown in Table.\ \ref{table:owcdm}. The measurements are 
similar to using the whole sample (z=0.16-0.44) but the constraints are worse. 
It might be due to some tension between the measurements of these two subsample 
which might be caused by higher noise level from smaller samples or by the evolution 
of the dark energy which would need a more complex model than owCDM. We will 
explore this issue in our future research.

\subsection{Constraints on owCDM Model} \label{sec:model-1D}

We now present the cosmological parameter constraints
for the owCDM model (non-flat Universe with a constant dark energy equation of state).
Table.\ \ref{table:owcdm} also shows the constraints from cosmological microwave 
background (WMAP7) and supernova (Union2 compilation) data and their combination
with SDSS LRG data. To include the constraints from WMAP7 \citep{Komatsu:2010fb}, 
we use the constraints on the CMB shift parameters $\{R,l_a\}$ 
and $z_*$ by \cite{Wang11} (see Appendix\ \ref{sec:cmb}).
To calculate the constraints from Union2 
SNe, we use the add-on code for cosmoMC which can be download from the 
website of Union2 SNe\footnote{http://supernova.lbl.gov/Union/}. 
For a given model, one could obtain $\chi^2$ for each data set, 
i.e. $\chi^2_{CMB}$ and $\chi^2_{SN}$\footnote{While computing 
$\chi^2_{SN}$, we use the covariance matrix with systematics to 
obtain more reliable constraints from SNe}. To include the constraint 
we obtained from the galaxy clustering data, one should add the following 
term to the $\chi^2$ with

\begin{equation} \label{eq:chi2_1d}
 \chi^2_{LRG} = \left[\frac{r_s(z_d)/D_V(0.35)-0.1143}{0.0033}\right]^2
\end{equation}

\begin{figure}
\centering
\includegraphics[width=0.8\linewidth,clip]{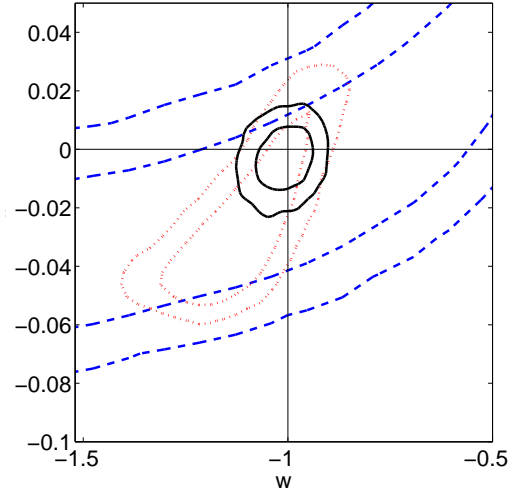}
\caption{2D marginalized
  contours for $68\%$ and $95\%$ for $w$ and $\Omega_k$ (owCDM model assumed)
  from WMAP7 (dashed blue), WMAP7+Union2 SN (dotted red),
  and WMAP7+Union2 SN+LRG (solid black). The straight
  solid black lines indicate that $w=-1$ and $\Omega_k=0$.}
  \label{fig:w_vs_ok}
\end{figure}

Combining all three data sets, LRG, CMB, and SNe, and assuming the owCDM model, 
we find that $\Omega_k=-0.0032^{+0.0074}_{-0.0072}$, and $w=-1.010^{+0.046}_{-0.045}$, 
which is consistent with $\Lambda$CDM model (in agreement with previous work, see e.g.,
\citealt{Serra09,Wang09,Mortonson10,Zhao10}).
Fig.\ \ref{fig:w_vs_ok} compares the constraints on $w$ and $\Omega_k$ in the
owCDM model. We can see that the addition of SDSS LRG data significantly tightens the constraints on
dark energy and cosmological parameters.

\begin{table*}
\begin{center}\scriptsize
\begin{tabular}{c}
\\
owCDM \\
\end{tabular}

\begin{tabular}{ccccccc}\hline
       &$\Omega_m$  &$\Omega_X$&w &$H_0$ &$\Omega_mh^2$&$\Omega_k$ \\ \hline
CMB										
&\ \ $	0.325	_{	-0.077	}^{+	0.072	}$
&\ \ $	0.67	_{	-0.071	}^{+	0.073	}$
&\ \ $	-1.04	_{	-0.63	}^{+	0.62	}$
&\ \ $	65.7	_{	-7.2	}^{+	8.2	}$
&\ \ $	0.1356	_{	-0.0060	}^{+	0.0059	}$
&\ \ $	0.006	_{	-0.048	}^{+	0.058	}$
\\
CMB+SN						
&\ \ $	0.348	_{	-0.072	}^{+	0.06	}$
&\ \ $	0.676	_{	-0.045	}^{+	0.052	}$
&\ \ $	-1.11	\pm 0.11$
&\ \ $	63.4	_{	-5.8	}^{+	6.5	}$
&\ \ $	0.1361	_{	-0.0060	}^{+	0.0061	}$
&\ \ $	-0.025	_{	-0.019	}^{+	0.022	}$
\\
CMB+SN+LRG					
&\ \ $	0.279	_{	-0.019	}^{+	0.018	}$
&\ \ $	0.725	\pm 0.018$
&\ \ $	-1.01	_{	-0.045	}^{+	0.046	}$
&\ \ $	69.9	\pm 2.3$
&\ \ $	0.1357	_{	-0.0059	}^{+	0.0061	}$
&\ \ $	-0.0032	_{	-0.0072	}^{+	0.0074	}$
\\				
CMB+SN+LRG2Z								
&\ \ $	0.278	\pm 0.032$
&\ \ $	0.725	\pm 0.026$
&\ \ $	-1.008	_{	-0.054	}^{+	0.053	}$
&\ \ $	70.3	\pm 4.2$
&\ \ $	0.136	_{	-0.0063	}^{+	0.0061	}$
&\ \ $	-0.003	\pm 0.011$
\\ \hline
\end{tabular}

\end{center}
\caption{Constraints of the cosmological parameters from various
data combinations with owCDM model assumed, where LRG is using the fiducial result of this paper (eq.\ \ref{eq:chi2_1d})
and LRG2Z is using the measurements from two subsamples as described in Sec.\ \ref{sec:2slice}. There are two inferred parameters, $\Omega_m$
and $\Omega_k$, in this table.} 
\label{table:owcdm}
\end{table*}

\subsection{Validation Using Mock Catalogs}
\label{sec:lasdamas_result}

In order to validate our method, we have applied it to the 2D 2PCF 
of 40 LasDamas mock catalogs (which are indexed with 01a-40a). Again, we 
apply the flat and wide priors ($\pm7\sigma_{WMAP7}$) on $\Omega_b h^2$ 
and $n_s$, centered on the input values of the simulation 
($\Omega_b h^2=0.0196$ and $n_s=1$).

Table \ref{table:mean_lasdamas_full} shows our measurements of
$\{D_V(0.35)$, $\Omega_m h^2$, $r_s(z_d)/D_V(0.35)$, $A(0.35)\}$ from the LasDamas mock catalogs of
the SDSS LRG sample. These are consistent with the input
parameters, establishing the validity of our method.

\begin{table}
\begin{center}
\begin{tabular}{crr|r}\hline
&mean &$\sigma$ &input value\\ \hline
 $D_V(0.35)$(Mpc)&\ \ 1349 &\ \ 69 &\ \ 1356 \\
 $\Omega_m h^2$&\ \ 0.120 &\ \ 0.015 &\ \  0.1225 \\
 $r_s(z_d)/D_V(0.35)$&\ \ 0.1205 &\ \ 0.0059  &\ \ 0.1175 \\
 $A(0.35)$&\ \ 0.441 &\ \ 0.026 &\ \  0.452 \\
\hline
\end{tabular}
\end{center}
\caption{The mean, standard deviation, and the 68\% C.L. bounds of
$\{D_V$(0.26), $\Omega_m h^2$, $r_s(z_d)/D_V(0.35)$, $A(0.35)\}$ from the LRGfull mock catalogs
of the LasDamas simulations. Our measurements are consistent with the
input values within 1$\sigma$.
} \label{table:mean_lasdamas_full}
\end{table}

\subsection{Comparison with other analyses} \label{sec:comparison}

There have been several analyses of the clustering of the
SDSS LRG spectroscopic sample. \cite{Percival:2009xn} use power
spectrum analysis to extract the BAO signals. The sample they use
include not only LRGs but also main galaxies sample. They obtain
the constraint on $r_s/D_V(0.35) = 0.1097\pm0.0036$. 
\cite{Reid:2009xm} apply power spectrum analysis on the
reconstructed halo density field derived from SDSS DR7full
(flux-limited LRG) sample. Their measurement of 
$r_s/D_V(0.35) = 0.1097^{+0.0039}_{-0.0042}\, (k_{max}=0.2)$ which 
is similar to \cite{Percival:2009xn}. 
There is about $1\sigma$ difference between their results and ours.
However, in \cite{Reid:2009xm}, they show a dependence on the 
minimum scales of the range, i.e. $r_s/D_V(0.35) = 0.1118^{+0.0043}_{-0.0046}\, 
(k_{max}=0.15)$ and $r_s/D_V(0.35) = 0.1136^{+0.0070}_{-0.0072}\, (k_{max}=0.1)$. 
Their results are more consistent with ours when a smaller $k_{max}$
(i.e., a larger minimum scale) is used, which represents a
more conservative choice for the scale range.

\section{Systematic Tests} \label{sec:test}

Table.\ \ref{table:test} shows the systematic tests that we have
done varying key assumptions made in our analysis.
These include the range of scales used to calculate the correlation function, 
the nonlinear damping scale, an overall shift 
in the measured correlation function due to a systematic error. 


We vary the effective redshift (from $z_{eff}=0.33$ to $z_{eff}=0.35$) used 
to calculate the theoretical model. We rescale the results to $z=0.35$ for 
comparison and find that the results are insensitive to the effective redshift.

We also test the sensitivity of our results to the nonlinear damping scale, 
$k_{\star}$. Although $k_{\star}$ can be predicted 
accurately in the real space \citep{Crocce:2005xz,Matsubara:2007wj}, 
in the redshift space, it would also depend on the redshift distortions 
which cannot be well determined from the spherically-averaged correlation 
function. In table.\ \ref{table:test}, one can tell that the results are 
not sensitive to $k_{\star}$.

In principle, the range of scales chosen for the analysis should be
as large as possible, in order to derive the tightest constraints. 
However, we do not use the small scales ($s<40\ h^{-1}$Mpc),
where the scale dependence of redshift distortion and galaxy bias are not
negligible and cannot be accurately determined at present. 
According to Fig. 5 in \cite{Eisenstein:2005su}, these effects are
negligible at $s>40\ h^{-1}$Mpc. We vary the minimum
scales used and find that the $r_s(z_d)/D_V(0.35)$ is insensitive 
to it but $\Omega_m h^2$ is not. This indicates the robustness
of the measurement of $r_s(z_d)/D_V(0.35)$ (but not $\Omega_m h^2$) 
from this paper.

On larger scales ($s>130\ h^{-1}$Mpc), the observed correlation 
is significantly higher than expected in conventional models
of galaxy clustering. This high tail problem 
was reported in previous work, see, e.g., \cite{Eisenstein:2005su}, 
\cite{Hutsi:2005qv}, and \cite{Sanchez:2009jq}. They found that the observed 
correlation function could be fitted better by lowering all the data points 
by a constant. In other words, they assumed a constant shift from some 
systematic error. Although this systematic error is unknown, we could 
minimize its effect by using smaller scale. The reason is that the correlation 
function has larger value at smaller scale so that the results are less sensitive 
to the shift. We choose $s=120 h^{-1}Mpc$ as our boundary for the large scale 
and show that the results are insensitive to the constant shift by lowering down 
the data points of the observed correlation function by 0.002. 
We find that $\Omega_m h^2$ varies by 1$\sigma$ and $r_s/D_V(0.35)$ 
only varies by 0.2$\sigma$. Therefore, our measurement of $r_s/D_V(0.35)$ is 
robust to the systematic shift.  This is another indication that
our measurement of $r_s(z_d)/D_V(0.35)$ (but not that of $\Omega_m h^2$) 
is robust.

\begin{table*}
\begin{center}
\begin{tabular}{crrrr}\hline
    &$D_V(0.35)$(Mpc)&    $\Omega_mh^2$ &   $r_s(z_d)/D_V(0.35)$ &$A(0.35)$\\ \hline
fiducial model						
&\ \ $	1428	_{	-73	}^{+	74	}$
&\ \ $	0.105	\pm0.016	$
&\ \ $	0.1143	_{	-0.0031	}^{+	0.0029	}$
&\ \ $	0.439	\pm0.02$
\\
$z_{eff}=0.35$
&\ \ $	1427	_{	-71	}^{+	72	}$
&\ \ $	0.105	_{	-0.015	}^{+	0.016	}$
&\ \ $	0.1144	_{	-0.0031	}^{+	0.0030	}$
&\ \ $	0.439	\pm 0.02$
\\
$k_\star=0.11$				
&\ \ $	1426	_{	-72	}^{+	73	}$
&\ \ $	0.106	\pm0.015$
&\ \ $	0.1144	_{	-0.0031	}^{+	0.0030	}$
&\ \ $	0.439	\pm0.02$
\\				
s=20-120						
&\ \ $	1398	_{	-62	}^{+	64	}$
&\ \ $	0.116	\pm0.012$
&\ \ $	0.1139	\pm0.0028$
&\ \ $	0.453	\pm0.015$
\\
s=60-120					
&\ \ $	1418	_{	-93	}^{+	94	}$
&\ \ $	0.108	\pm0.025$
&\ \ $	0.1149	_{	-0.0033	}^{+	0.0030	}$
&\ \ $	0.438	\pm0.03$
\\
s=40-100					
&\ \ $	1448	_{	-89	}^{+	96	}$
&\ \ $	0.114	_{	-0.019	}^{+	0.018	}$
&\ \ $	0.111	_{	-0.0058	}^{+	0.0050	}$
&\ \ $	0.463	_{	-0.032	}^{+	0.033	}$
\\
s=40-140					
&\ \ $	1393	_{	-87	}^{+	90	}$
&\ \ $	0.110	\pm0.018$
&\ \ $	0.1164	_{	-0.0047	}^{+	0.0040	}$
&\ \ $	0.437	_{	-0.023	}^{+	0.025	}$
\\
shift = 0.002					
&\ \ $	1388	_{	-77	}^{+	80	}$
&\ \ $	0.122	\pm0.020$
&\ \ $	0.1136	_{	-0.0035	}^{+	0.0032	}$
&\ \ $	0.459	\pm0.023$
\\
\hline
\end{tabular}
\end{center}
\caption{This table shows the systematic tests with the scale range, 
the fiducial model used, the effective redshift, the damping factor, 
and the shift from a systematic error. The
fiducial results is obtained by assuming $\Lambda$CDM model with $\Omega_m=0.25$
as fiducial model considering the scale range ($s=40-120\ h^{-1}$Mpc), using the effective redshift ($z_{eff}=0.33)$), and 
the damping factor, $k_{\star}$, marginalized over with the a flat prior ($0.09<k_\star<0.13\ h$Mpc$^{-1}$).
The other results are calculated with only one
quantity different from the fiducial one. $n_s=0.963$ and $\Omega_bh^2=0.02258$ 
are marginalized with the same flat priors ($\pm 7\sigma_{WMAP7}$) in this paper.}
 \label{table:test}
\end{table*}

\section{Conclusion} \label{sec:conclusion}

We have presented our first results for the model independent constraints
on dark energy from the spherically-averaged correlation function of SDSS
DR7 data, using an MCMC likelihood analysis.
Our constraints on $\{D_V(0.35)$, $\Omega_mh^2$, $r_s(z_d)/D_V(0.35)$, $A(0.35)\}$
are summarized by Table\ \ref{table:mean} and \ref{table:covar_matrix}.
Applying these results to constraining a constant dark energy equation
of state without assuming a flat Universe (the owCDM model), and combining
with WMAP7 and Union2 SN data sets, we find that 
 $\Omega_k=-0.0032^{+0.0074}_{-0.0072}$ and $w=-1.010^{+0.046}_{-0.045}$, 
consistent with a flat universe with a cosmological constant ($\Omega_k=0$, $w=-1$).

We have also measured the model independent constraints of $H(0.35)$ and
$D_A(0.35)$ from the spherically-averaged correlation function from SDSS
DR7 LRGs, as a baseline for comparison with constraints from studies of the
2D correlation function (see \cite{Chuang:2011fy}). We find that 
$\{H(0.35)$, $D_A(0.35)\}$ from the spherically-averaged correlation function 
provide much weaker constraints than 
the 2D correlation function; this is as expected since 
spherically-averaging reduces the amount of information extractable from data.

The correlation function analysis is expected to be a more robust way to extract the 
BAO signals than the power spectrum analysis, because one can easily get rid 
of the systematic uncertainties such as the redshift distortion, the galaxy bias,
and the non-linear effect by cutting off the small scale range \citep{Sanchez:2008iw}.

The power of the correlation function analysis is limited at present by
the available data. The correlation funciton that we have measured from the
SDSS DR7 data has a high tail (larger than expected correlations) at large scales
($s>120$) (see Fig.~\ref{fig:kazin_my}). Whether this high tail is simply
due to the sample variance or some other systematic issue, e.g., extinction 
correction, will only become clear as more ambitious galaxy survey data become
available in the future (e.g., from BOSS\footnote{http://www.sdss3.org/cosmology.php},
or Euclid\footnote{See http://sci.esa.int/euclid, and
\cite{Cimatti09,Laureijs09,Wang10}}).

\section*{Acknowledgements}

We would like to thank Michael Blanton, Daniel Eisenstein, Alex Kim, Antony Lewis, 
Ariel Sanchez, and Martin White for useful comments. We are grateful to the 
LasDamas project for making their mock catalogs publicly available. The computing 
for this project was performed at the OU Supercomputing Center for Education and 
Research (OSCER) at the University of Oklahoma (OU). OSCER Director Henry Neeman 
and HPC Application Software Specialist
Joshua Alexander provided invaluable technical support.
This work was supported in part by DOE grant DE-FG02-04ER41305.

\setlength{\bibhang}{2.0em}

\appendix
\section{LN mock catalogs} \label{sec:lognormal}

One convenient way to generate mock galaxy catalogs for
calculating covariance matrix is using lognormal random fields
which can approximate the present-day non-linear fluctuation
field\ \citep{Coles:1991if}.
We create 500 lognormal (LN) density
fields \citep{Coles:1991if,Percival:2003pi} on a $512^3$ grid with
box length $4096\ h^{-1}$Mpc. We then draw a random Poisson
variable with mean given by the selection functions and lognormal
field to create the mock catalogs. We follow the steps described in
\cite{Percival:2003pi} except that we don't cut the input power at
0.25 Nyquist frequency because it makes the restored correlation
function deviate from the input one. With a input correlation function,
$\xi(r)$, the Gaussian field correlation function is obtained by
\begin{equation}
\xi_G(r)=\ln[1+\xi(r)],
\end{equation}
and this can be Fourier transformed to the power spectrum, $P_G(k)$. A Gaussian density
field $\delta_G(r)$ is generated on the grid with this power
spectrum, and the corresponding lognormal field is calculated by
\begin{equation}
\delta_{LN}(r)=\exp\left[\delta_G(r)-\frac{\sigma_G^2}{2}\right]-1,
\end{equation}
where $1+\delta_{LN}(r)$ is the lognormal density field which is always positive
by definition and $\sigma_G^2$ is the variance of the  Gaussian density field which
can be calculated by
\begin{equation}
 \sigma_G^2=\sum^{N_{grid}}_{i,j,l=1}P_G\left[(k^2_{x_i}+k^2_{y_j}
 +k^2_{z_l})^\frac{1}{2}\right],
\end{equation}
where $N_{grid}$ is the number of grid points, $k_{m_n}=\frac{2\pi}{L}\left(n-\frac{N_{grid}}{2}\right)$, $L$ is the box length, and $m=x$, $y$, or $z$. Then, the mock
catalogs can be constructed by drawing the Poisson random variables with the
means given by this lognormal field and the selection function of the galaxy survey.

To compute the correlation function of these mock catalogs, one should create the random data
on the same grid as well to cancel out the effect of the finite
size of the grid.  The input correlation function in this study is the
theoretical correlation function with parameters ($\Omega_m=0.25$,
$\Omega_b=0.04, h=0.7, n_s=1$) which are the same as the input parameters of the LasDamas simulations. We fix $k^\star=0.11$ and the amplitude is adjusted to fit the averaged correlation function from the LasDamas mock catalogs we use. We are not fiting the observed correlation function because we want to find out is whether the LN mock catalogs could behave as good as LasDamas mock catalogs while estimating the covariance matrix.

\section{Normality test}
\label{sec:normality_test}

\begin{figure}
\centering
\includegraphics[width=0.8\linewidth,clip,angle=-90]{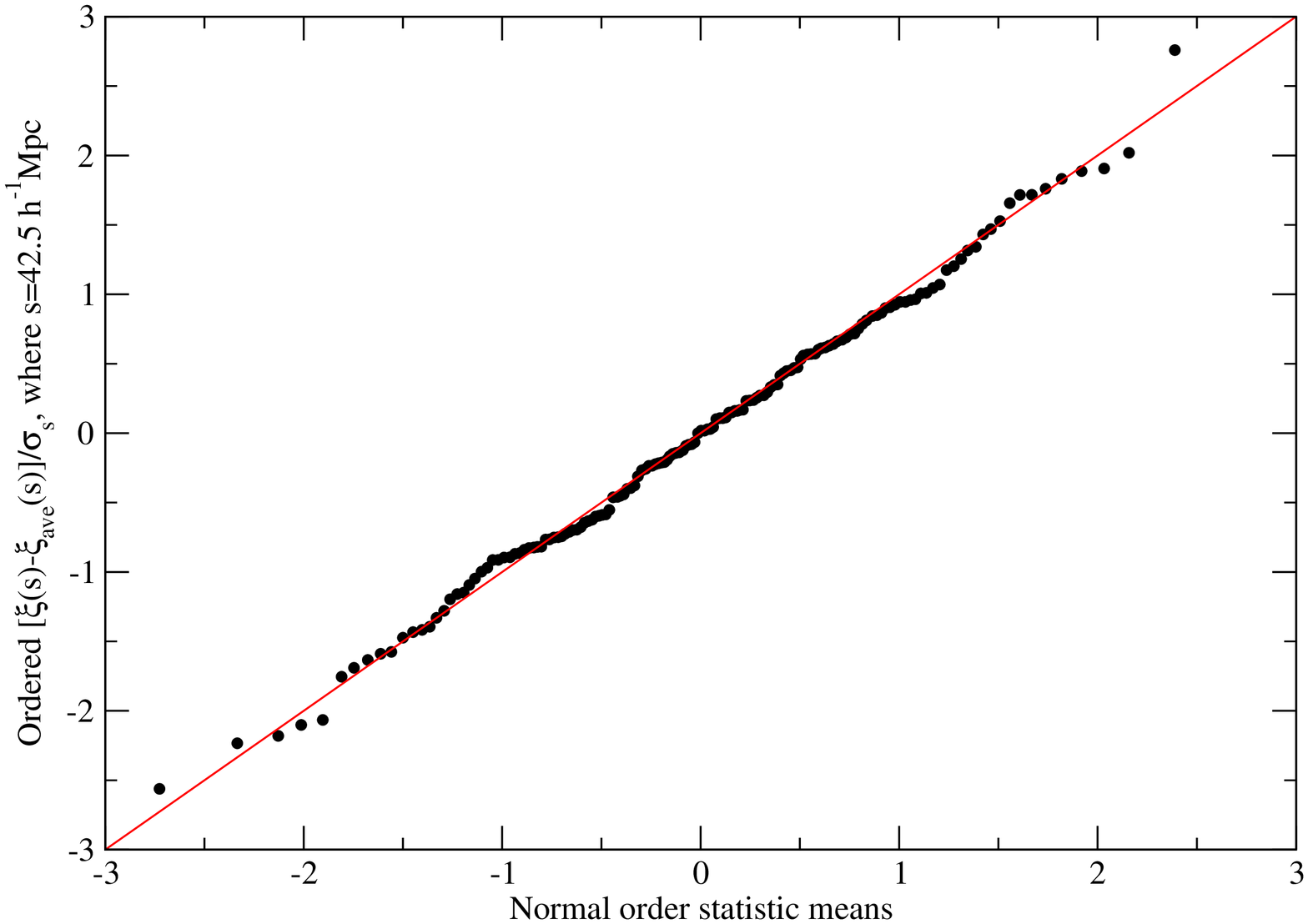}
\caption{The normality probability plot of the first bins, $40<s<45 h^{-1}$Mpc, of correlation functions from the LasDamas mock catalogs we use for estimating the covariance matrix. The vertical axis is from the ordered values of the first bins from 160 mock catalogs which have been shifted and linear rescaled to have zero mean and unity variance. The horizontal axis is from the expected values of 160 ordered gaussian random numbers with zero mean and unity variance. That it is an approximate straight line means the bin values are approximately normally distributed.}
  \label{fig:normality_test1}
\end{figure}

\begin{figure}
\centering
\includegraphics[width=0.8\linewidth,clip,angle=-90]{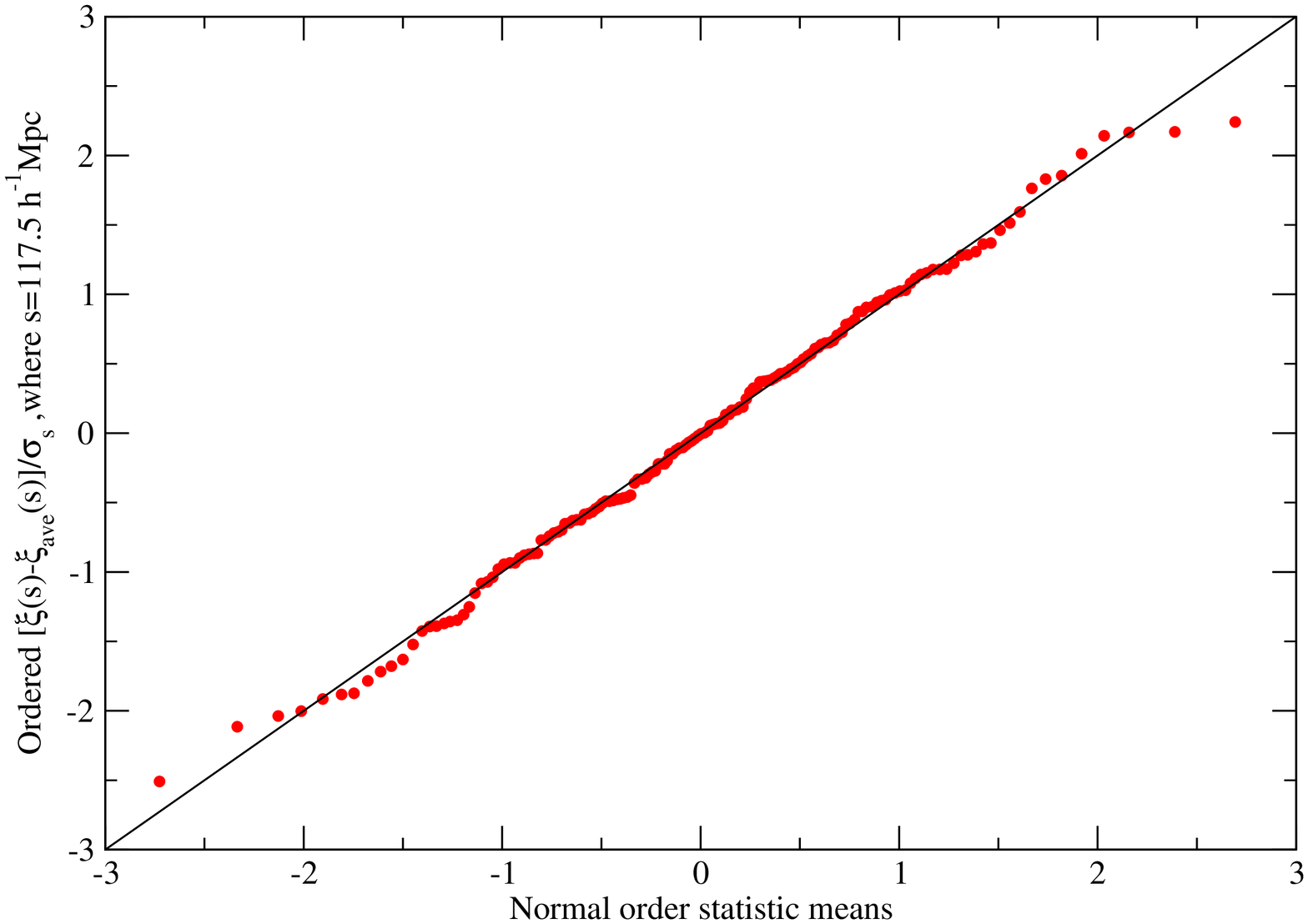}
\caption{The normality probability plot of the first bins, $115<s<120 h^{-1}$Mpc, of correlation functions from the LasDamas mock catalogs we use for estimating the covariance matrix. The vertical axis is from the ordered values of the first bins from 160 mock catalogs which have been shifted and linear rescaled to have zero mean and unity variance. The horizontal axis is from the expected values of 160 ordered gaussian random numbers with zero mean unity variance. That it is an approximate straight line means the bin values are approximately normally distributed.}
  \label{fig:normality_test2}
\end{figure}

We check the normality of the correlation functions from the LasDamas 
mock catalogs by showing the normal probability plots of the first and 
last bin we use (see Fig.\ \ref{fig:normality_test1} and 
\ref{fig:normality_test2}). One can tell that they are well 
described by a normal distribution.


\section{CMB distance priors} \label{sec:cmb}
\cite{Wang07} showed that CMB shift parameters $(l_a, R)$, together with
$\Omega_b h^2$, provide an efficient and intuitive summary of CMB data as far as 
dark energy constraints are concerned. It is equivalent to replace
$\Omega_b h^2$ with $z_*$, the redshift to the photon-decoupling surface
\citep{Wang09}.

The CMB shift parameters are defined as \citep{Wang07}:
\ba
R \equiv \sqrt{\Omega_m H_0^2} \,r(z_*),
l_a \equiv \pi r(z_*)/r_s(z_*),
\ea 
and $z_{\star}$ is the redshift to the photon-decoupling surface given by 
the fitting formula\citep{Hu:1995en}:
\be
z_*=1048\, \left[1+ 0.00124 (\Omega_b h^2)^{-0.738}\right]\,
\left[1+g_1 (\Omega_m h^2)^{g_2} \right],
\ee
where
\ba
g_1 &= &\frac{0.0783\, (\Omega_b h^2)^{-0.238}}
{1+39.5\, (\Omega_b h^2)^{0.763}}\\
g_2 &= &\frac{0.560}{1+21.1\, (\Omega_b h^2)^{1.81}}
\ea
The comoving distance to an object at redshift $z$ is given by:
\ba
\label{eq:r(z)}
 & &r(z)=cH_0^{-1}\, |\Omega_k|^{-1/2} {\rm sinn}[|\Omega_k|^{1/2}\, \Gamma(z)],\\
 & &\Gamma(z)=\int_0^z\frac{dz'}{E(z')}, \hskip 1cm E(z)=H(z)/H_0 \nonumber
\ea
where ${\rm sinn}(x)=\sin(x)$, $x$, $\sinh(x)$ for 
$\Omega_k<0$, $\Omega_k=0$, and $\Omega_k>0$ respectively;
and the expansion rate the universe $H(z)$ is given by
\ba
\label{eq:H(z)}
&&H^2(z)  \equiv  \left(\frac{\dot{a}}{a}\right)^2 \\
 &= &H_0^2 \left[ \Omega_m (1+z)^3 +\Omega_r (1+z)^4 +\Omega_k (1+z)^2 
+ \Omega_X X(z) \right],\nonumber
\ea
where $\Omega_m+\Omega_r+\Omega_k+\Omega_X=1$, and
the dark energy density function $X(z)$ is defined as
\be
X(z) \equiv \frac{\rho_X(z)}{\rho_X(0)}.
\ee
Note that $\Omega_r \ll \Omega_m$, thus the $\Omega_r$ term
is usually omitted in dark energy studies, since dark energy
should only be important at late times.

The comoving sound horizon at redshift $z$ is given by
\ba
\label{eq:rs}
r_s(z)  &= & \int_0^{t} \frac{c_s\, dt'}{a}
=cH_0^{-1}\int_{z}^{\infty} dz'\,
\frac{c_s}{E(z')}, \nonumber\\
 &= & cH_0^{-1} \int_0^{a} 
\frac{da'}{\sqrt{ 3(1+ \overline{R_b}\,a')\, {a'}^4 E^2(z')}},
\ea
where $a$ is the cosmic scale factor, $a =1/(1+z)$, and
$a^4 E^2(z)=\Omega_m (a+a_{\rm eq})+\Omega_k a^2 +\Omega_X X(z) a^4$,
with $a_{\rm eq}=\Omega_{\rm rad}/\Omega_m=1/(1+z_{\rm eq})$, and
$z_{\rm eq}=2.5\times 10^4 \Omega_m h^2 (T_{CMB}/2.7\,{\rm K})^{-4}$.
The sound speed is $c_s=1/\sqrt{3(1+\overline{R_b}\,a)}$,
with $\overline{R_b}\,a=3\rho_b/(4\rho_\gamma)$,
$\overline{R_b}=31500\Omega_bh^2(T_{CMB}/2.7\,{\rm K})^{-4}$.
We take $T_{CMB}=2.725$.

The redshift of the drag epoch $z_d$ is well approximated by 
\cite{Eisenstein:1997ik}
\begin{equation}
z_d  =
 \frac{1291(\Omega_mh^2)^{0.251}}{1+0.659(\Omega_mh^2)^{0.828}}
\left[1+b_1(\Omega_bh^2)^{b2}\right],
\label{eq:zd}
\end{equation}
where
\begin{eqnarray}
  b_1 &= &0.313(\Omega_mh^2)^{-0.419}\left[1+0.607(\Omega_mh^2)^{0.674}\right],\\
  b_2 &= &0.238(\Omega_mh^2)^{0.223}.
\end{eqnarray}

There are only four independent parameters among these five and $n_s$ is marginalized over in this study. Therefore, there are only three parameters left, $\{l_a, R, z_*\}$.
CMB data are included in our analysis by adding
the following term to the $\chi^2$ of a given model
with $\Delta p_1=l_a(z_*)-302.35$, $\Delta p_2=R(z_*)-1.728$, and $\Delta p_3=z_*-1091.32$:
\be
\label{eq:chi2CMB}
\chi^2_{CMB}=\Delta p_i \left[ \mbox{Cov}^{-1}_{CMB}(p_i,p_j)\right]
\Delta p_j,
\ee 
where the inverse covariance matrix of $(l_a, R, z_*)$ from 
WMAP7 \citep{Komatsu:2010fb} is given by \citep{Wang11}:
\be
\mbox{Cov}_{CMB}^{-1}=
\left(
\begin{array}{ccc}
   1.85710  & 25.9289 & -1.14325\\
   25.9289  & 5963.26 & -99.3185\\
  -1.14325  &-99.3185  & 2.94429
\label{eq:CMB_cov}
\end{array}
\right)
\ee  

\label{lastpage}


\begin{thebibliography}{}

  \setlength{\itemindent}{-2.5em}

\bibitem[Abazajian et al.(2009)]{Abazajian:2008wr}
  Abazajian, K.~N., {\it et al.}  [SDSS Collaboration],
  Astrophys.\ J.\ Suppl.\  {\bf 182}, 543 (2009)
  [arXiv:0812.0649 [astro-ph]].

\bibitem[Amanullah et al.(2010)]{Amanullah:2010vv}
  Amanullah, R., {\it et al.},
  Astrophys.\ J.\  {\bf 716}, 712 (2010)
  [arXiv:1004.1711 [astro-ph.CO]].

\bibitem[Blake et al.(2007)]{Blake:2006kv}
  Blake, C.; Collister, A.; Bridle, S.; and Lahav,
O.,
  Mon.\ Not.\ Roy.\ Astron.\ Soc.\  {\bf 374}, 1527 (2007)
  [arXiv:astro-ph/0605303].

\bibitem[Blanton et al.(2005)]{Blanton:2004aa}
  Blanton, M.~R., {\it et al.}  [SDSS Collaboration],
  Astron.\ J.\  {\bf 129}, 2562 (2005)
  [arXiv:astro-ph/0410166].


\bibitem[Blanton \& Roweis(2007)]{Blanton:2006kt}
  Blanton, M.~R.; and Roweis,
S.,
  Astron.\ J.\  {\bf 133}, 734 (2007)
  [arXiv:astro-ph/0606170].

\bibitem[Cabre \& Gaztanaga(2008)]{Cabre:2008sz}
  Cabre, A.; and Gaztanaga, E.,
  arXiv:0807.2460 [astro-ph].

\bibitem[Chuang \& Wang(2011)]{Chuang:2011fy}
  C.~H.~Chuang and Y.~Wang,
  arXiv:1102.2251 [astro-ph.CO].

\bibitem[Cimatti et al.(2009)]{Cimatti09}
Cimatti, A.; Robberto, M.; Baugh, C.; Beckwith, S. V. W.; Content, R.;
Daddi, E.; De Lucia, G.; Garilli, B.; Guzzo, L.; Kauffmann, G.;
Lehnert, M.; Maccagni, D.; Martínez-Sansigre, A.; Pasian, F.; Reid, I. N.;
Rosati, P.; Salvaterra, R.; Stiavelli, M.; Wang, Y.; Osorio, M. Zapatero;
the SPACE team, Experimental Astronomy, 23, 39 (2009)

\bibitem[Coles \& Jones(1991)]{Coles:1991if}
  Coles, P., and Jones,
B.,
  Mon.\ Not.\ Roy.\ Astron.\ Soc.\  {\bf 248} (1991) 1.



\bibitem[Crocce \& Scoccimarro(2006)]{Crocce:2005xz}
  M.~Crocce and R.~Scoccimarro,
  Phys.\ Rev.\  D {\bf 73}, 063520 (2006)
  [arXiv:astro-ph/0509419].

\bibitem[Eisenstein \& Hu(1998)]{Eisenstein:1997ik}
  Eisenstein, D.~J.; and Hu,
W.,
  Astrophys.\ J.\  {\bf 496}, 605 (1998)
  [arXiv:astro-ph/9709112].

\bibitem[Eisenstein et al.(2001)]{Eisenstein:2001cq}
 Eisenstein, D.~J., {\it et al.}  [SDSS Collaboration],
  Astron.\ J.\  {\bf 122}, 2267 (2001)
  [arXiv:astro-ph/0108153].

\bibitem[Eisenstein et al.(2005)]{Eisenstein:2005su}
  Eisenstein, D.~J., {\it et al.}  [SDSS Collaboration],
  Astrophys.\ J.\  {\bf 633}, 560 (2005)
  [arXiv:astro-ph/0501171].

\bibitem[Eisenstein, Seo, and White(2007)]{Eisenstein:2006nj}
  Eisenstein, D.~J.; Seo, H.~j.; and White,
M.~J.,
  Astrophys.\ J.\  {\bf 664}, 660 (2007)
  [arXiv:astro-ph/0604361].

\bibitem[Fioc \& Rocca-Volmerange(1997)]{Fioc:1997sr}
  Fioc, M.; and Rocca-Volmerange,
B.,
  Astron.\ Astrophys.\  {\bf 326}, 950 (1997)
  [arXiv:astro-ph/9707017].

\bibitem[Fukugita et al.(1996)]{Fukugita:1996qt}
  Fukugita, M.; Ichikawa, T.; Gunn, J.~E.; Doi, M.; Shimasaku, K.; and Schneider, D.~P.,
  Astron.\ J.\  {\bf 111}, 1748 (1996).

\bibitem[Gunn et al.(1998)]{Gunn:1998vh}
  Gunn, J.~E., {\it et al.}  [SDSS Collaboration],
  Astron.\ J.\  {\bf 116}, 3040 (1998)
  [arXiv:astro-ph/9809085].
\bibitem[Gunn et al.(2006)]{Gunn:2006tw}
  Gunn, J.~E., {\it et al.}  [SDSS Collaboration],
  Astron.\ J.\  {\bf 131}, 2332 (2006)
  [arXiv:astro-ph/0602326].

\bibitem[Hemantha et al., in preparation]{Hemantha:2011}
  Hemantha, M. D. P.; Wang, Y.; and Chuang, C., in preparation

\bibitem[Hu and Sugiyama(1996)]{Hu:1995en}
  W.~Hu and N.~Sugiyama,
  Astrophys.\ J.\  {\bf 471}, 542 (1996)
  [arXiv:astro-ph/9510117].

\bibitem[Hutsi(2005)]{Hutsi:2005qv}
  Hutsi, G.,
  arXiv:astro-ph/0507678.

\bibitem[Kaiser(1987)]{Kaiser:1987qv}
  N.~Kaiser,
  Mon.\ Not.\ Roy.\ Astron.\ Soc.\  {\bf 227}, 1 (1987).

\bibitem[Kazin et al.(2010)]{Kazin:2009cj}
  Kazin, E.~A., {\it et al.},
  Astrophys.\ J.\  {\bf 710}, 1444 (2010)
  [arXiv:0908.2598 [astro-ph.CO]].

\bibitem[Komatsu et al.(2010)]{Komatsu:2010fb}
  Komatsu, E., {\it et al.},
  arXiv:1001.4538 [astro-ph.CO].

\bibitem[Landy \& Szalay(1993)]{Landy:1993yu}
  Landy, S.~D.; and Szalay,
A.~S.,
  Astrophys.\ J.\  {\bf 412}, 64 (1993).

\bibitem[Laureijs et al.(2009)]{Laureijs09}
Laureijs, R. et al. 2009,
``Euclid Assessment Study Report for the ESA Cosmic Visions'',
arXiv:0912.0914

\bibitem[Lewis, Challinor, \& Lasenby(2000)]{Lewis:1999bs}
  Lewis, A.; Challinor, A.; and Lasenby,
A.,
  Astrophys.\ J.\  {\bf 538}, 473 (2000)
  [arXiv:astro-ph/9911177].

\bibitem[Lewis \& Bridle(2002)]{Lewis:2002ah}
  Lewis, A., and Bridle,
S.,
  Phys.\ Rev.\  D {\bf 66}, 103511 (2002)
  [arXiv:astro-ph/0205436].

\bibitem[Martinez et al.(2009)]{Martinez:2008iu}
  Martinez, V.~J., {\it et al.},
  Astrophys.\ J.\  {\bf 696}, L93 (2009)
  [Erratum-ibid.\  {\bf 703}, L184 (2009)]
  [Astrophys.\ J.\  {\bf 703}, L184 (2009)]
  [arXiv:0812.2154 [astro-ph]].

\bibitem[Matsubara(2007)]{Matsubara:2007wj}
  T.~Matsubara,
  Phys.\ Rev.\  D {\bf 77}, 063530 (2008)
  [arXiv:0711.2521 [astro-ph]].

\bibitem[Mortonson, Hu, \& Huterer(2010)]{Mortonson10}
Mortonson, M.~J.; Hu, W.; Huterer, D., 2010, PRD, 81, 063007

\bibitem[Okumura et al.(2008)]{Okumura:2007br}
  Okumura, T.; Matsubara, T.; Eisenstein, D.~J.; Kayo, I.;
  Hikage, C.; Szalay, A.~S.; and Schneider,
D.~P.,
  Astrophys.\ J.\  {\bf 676}, 889 (2008)
  [arXiv:0711.3640 [astro-ph]].

\bibitem[Padmanabhan et al.(2007)]{Padmanabhan:2006ku}
  Padmanabhan, N., {\it et al.}  [SDSS Collaboration],
  Mon.\ Not.\ Roy.\ Astron.\ Soc.\  {\bf 378}, 852 (2007)
  [arXiv:astro-ph/0605302].

\bibitem[Penzias \& Wilson(1965)]{Penzias:1965wn}
  Penzias, A.~A.; and Wilson,
R.~W.,
  Astrophys.\ J.\  {\bf 142}, 419 (1965).

\bibitem[Percival, Verde, \& Peacock(2004)]{Percival:2003pi}
  Percival, W.~J.; Verde, L.; and Peacock,
J.~A.,
  Mon.\ Not.\ Roy.\ Astron.\ Soc.\  {\bf 347}, 645 (2004)
  [arXiv:astro-ph/0306511].

\bibitem[Percival et al.(2007)]{Percival:2007yw}
  Percival, W.~J.; Cole, S.; Eisenstein, D.~J.; Nichol, R.~C.;
  Peacock, J.~A.; Pope, A.~C.; and Szalay,
A.~S.,
  Mon.\ Not.\ Roy.\ Astron.\ Soc.\  {\bf 381}, 1053 (2007)
  [arXiv:0705.3323 [astro-ph]].

\bibitem[Percival et al.(2010)]{Percival:2009xn}
  Percival, W.~J., {\it et al.},
  Mon.\ Not.\ Roy.\ Astron.\ Soc.\  {\bf 401}, 2148 (2010)
  [arXiv:0907.1660 [astro-ph.CO]].

\bibitem[Perlmutter et al.(1999)]{Perlmutter:1998np}
  Perlmutter, S., {\it et al.}  [Supernova Cosmology Project Collaboration],
  Astrophys.\ J.\  {\bf 517}, 565 (1999)
  [arXiv:astro-ph/9812133].

\bibitem[\protect\citeauthoryear{Press et al.}{1992}]{press92}
    Press W.H., Teukolsky S,A., Vetterling W.T., Flannery B.P.,
    1992, Numerical recipes in C. The art of scientific computing,
    Second edition, Cambridge: University Press.

\bibitem[Reid et al.(2010)]{Reid:2009xm}
  Reid, B.~A., {\it et al.},
  2010, MNRAS, 404, 60

\bibitem[Riess et al.(1998)]{Riess:1998cb}
  Riess, A.~G., {\it et al.}  [Supernova Search Team Collaboration],
  Astron.\ J.\  {\bf 116}, 1009 (1998)
  [arXiv:astro-ph/9805201].

\bibitem[Sanchez, Baugh, and Angulo(2008)]{Sanchez:2008iw}
  Sanchez, A.~G.; Baugh, C.~M.; and Angulo,
R.,
  Mon.\ Not.\ Roy.\ Astron.\ Soc.\  {\bf 390}, 1470 (2008)
  [arXiv:0804.0233 [astro-ph]].

\bibitem[Sanchez et al.(2009)]{Sanchez:2009jq}
  Sanchez, A.~G.; Crocce, M.; Cabre, A.; Baugh, C.~M.; and Gaztanaga,
E.,
  arXiv:0901.2570 [astro-ph].


\bibitem[Seo \& Eisenstein(2003)]{SE03}
Seo, H., Eisenstein, D. J., 2003, ApJ, 598, 720

\bibitem[Serra et al.(2009)]{Serra09}
Serra, P.; Cooray, A.; Holz, D.~ E.; Melchiorri, A.; Pandolfi, S.; Sarkar, D.,
2009, PRD, 80, 121302


\bibitem[Smith et al.(2003)]{Smith:2002dz}
  Smith, R.~E., {\it et al.}  [The Virgo Consortium Collaboration],
  Mon.\ Not.\ Roy.\ Astron.\ Soc.\  {\bf 341}, 1311 (2003)
  [arXiv:astro-ph/0207664].

\bibitem[Strauss et al.(2002)]{Strauss:2002dj}
  Strauss, M.~A., {\it et al.}  [SDSS Collaboration],
  Astron.\ J.\  {\bf 124}, 1810 (2002)
  [arXiv:astro-ph/0206225].

\bibitem[Tegmark et al.(2004)]{Tegmark:2003uf}
  Tegmark, M., {\it et al.}  [SDSS Collaboration],
  Astrophys.\ J.\  {\bf 606}, 702 (2004)
  [arXiv:astro-ph/0310725].

\bibitem[Wang \& Mukherjee(2007)]{Wang07}	
Wang, Y.; Mukherjee, P. 2007, Phys.Rev.D, 76, 103533

\bibitem[Wang(2009)]{Wang09}
Wang, Y., 2009, PRD, 80, 123525

\bibitem[Wang et al.(2010)]{Wang10}
Wang, Y., et al., 2010, MNRAS, 409, 737 

\bibitem[Wang, Chuang, \& Mukherjee(2011)]{Wang11}	
Wang, Y.; Chuang, C.-H., \& Mukherjee, P. 2011, 
arXiv:1109.3172


\bibitem[Zehavi et al.(2005)]{Zehavi:2004zn}
  Zehavi, I., {\it et al.}  [SDSS Collaboration],
  Astrophys.\ J.\  {\bf 621}, 22 (2005)
  [arXiv:astro-ph/0411557].

\bibitem[Zhao \& Zhang(2010)]{Zhao10}
Zhao, G.-B.; Zhang, X., PRD, 81, 043518

\end{thebibliography}
\end{document}